\documentclass[aps,prc]{revtex4}
\usepackage{graphicx}
\usepackage{amsmath}
\begin{document}
\title{Analyzing Correlation Functions with Tesseral and Cartesian Spherical Harmonics}
\author{Pawel Danielewicz}
\affiliation{Department of Physics and Astronomy, and National Superconducting
Cyclotron Laboratory,\\ Michigan State University, East Lansing, Michigan, 48824}
\author{Scott Pratt}
\affiliation{Department of Physics and Astronomy,\\
Michigan State University, East Lansing, Michigan, 48824}
\date{\today}

\begin{abstract}
  The dependence of inter-particle correlations on the orientation of particle
  relative-momentum can yield unique information on the space-time features of
  emission in reactions with multiparticle final states.  In the present paper,
  the benefits of a representation and analysis of the three-dimensional
  correlation information in terms of surface spherical harmonics is presented.
  The harmonics include the standard complex tesseral harmonics and the real
  cartesian harmonics.  Mathematical properties of the lesser-known cartesian
  harmonics are illuminated.  The physical content of different angular
  harmonic components in a correlation is described.  The resolving power of
  different final-state effects with regarding to determining angular features
  of emission regions is investigated.  The considered final-state effects
  include identity interference and strong and Coulomb interactions.  The
  correlation analysis in terms of spherical harmonics is illustrated with the
  cases of gaussian and blast-wave sources for proton-charged meson and
  baryon-baryon pairs.
\end{abstract}


\maketitle

\section{Introduction}
\label{sec:intro}
Measurements of low relative-velocity correlations yield access to the size and
shape of relative emission sources in reactions with multiparticle final states
\cite{Heinz:1999rw, Bauer:1993wq, Wiedemann:1999qn, Tomasik:2002rx}.
This information plays an especially important role in in heavy-ion collisions.
The duration of emission, which is linked to the
nuclear equation of state, can be also
inferred from those measurements.  Perhaps the most startling finding, during
the first years of the Relativistic Heavy Ion Collider (RHIC) program, was the
disagreement between the features of experimentally determined emission sources
\cite{starhbt, phenixhbt, phoboshbt} and predictions of hydrodynamic models
\cite{rischkehydro, teaney, bassdumitrusoff, Kolb:2003dz} incorporating an
anticipated phase transition, which would yield a long emission duration.
A long emission duration would have resulted in an elongated shape
($R_{\rm out}/R_{\rm side}>>1$ in the usual RHIC nomenclature)
of the emission source for particles of a specific momentum.
In the literature, the most
exhaustively studied correlation has been that produced by interference for
identical mesons. For pure interference, the correlation expressed as a
function of the relative particle momentum is related to the relative emission
source through a simple Fourier transformation, principally yielding direct
access to the source shape.  In the measured charged-meson correlations, the
Coulomb effects competing with interference had been normally compensated for
with cumbersome corrections.  However, as we demonstrate, the correlations
induced by the Coulomb and further by the strong interactions can also provide
information on the source shapes.  For any type of correlation, the analysis is
facilitated by the use of spherical harmonics, either the standard complex
tesseral harmonics or the real cartesian harmonics \cite{applequist89}.  The harmonics allow for a straightforward faithful representation of the
three-dimensional correlation data, and also better facilitate isolating
and focusing on specific physical properties of the source.  Different source
parameterizations will be employed for illustration of our discussions.  For
example, the $\ell=2$ harmonic expansion coefficients, characterizing
quadrupole deformation, provide information on the ratio of radii $R_{\rm
  out}/R_{\rm side}$ in the common gaussian source representation.

The correlation, associated with the final-state effects at low relative
velocities within a subsystem of the particles $a$ and $b$, can be linked to a
source function ${\mathcal S}$ through the relation:
\begin{equation}
\label{eq:corrmaster}
\frac{{\rm d}^6N^{ab}/{\rm d}^3p_a{\rm d}^3p_b}
{({\rm d} N^a{\rm d}^3p_a) \,
({\rm d}N^b{\rm d}^3p_b)} =
1+{\mathcal R}({\pmb P},{\pmb q})= \int {\rm d}^3r \, |\phi^{(-)}({\pmb q},{\pmb r})|^2 \,
{\mathcal S}({\pmb P},{\pmb r}) \, .
\end{equation}
Here, ${\mathcal R}({\pmb P},{\pmb q})$ is the deviation of the l.h.s.\
correlation function from unity, ${\pmb P}$ is the total momentum of the pair
and ${\pmb q}$ is the relative momentum measured in the two-particle rest frame.
The factor $|\phi^{(-)}|^2$ represents the relative wavefunction squared, in
the two-particle rest frame, with asymptotic condition imposed on the outgoing
wave.  The square of the wavefunction is averaged over spins.  The function
${\mathcal S}({\pmb P},{\pmb r})$ represents the probability density that the
particles $a$ and $b$, moving with the same velocity ${\pmb V}$ corresponding to
the total momentum ${\pmb P}$, are emitted from the reacting system a
distance ${\pmb r}$ apart in their rest frame.  Since at large ${\pmb q}$, the
emission in multiparticle final states becomes uncorrelated, i.e.\ ${\mathcal
  R} \rightarrow 0$ while $|\phi^{(-)}|^2$ averages to 1 as a function of
${\pmb r}$, it follows that ${\mathcal S}$ is normalized to 1, $\int {\rm d}^3 r
\, {\mathcal S}({\pmb P},{\pmb r}) = 1$.  A detailed derivation of Eq.\
(\ref{eq:corrmaster}) can be found in \cite{Anchishkin:1997tb}
(see also~\cite{Danielewicz:1992sh}).

The probability density of emission at the relative distance may be expressed
in terms of single-particle probability densities $s$:
\begin{equation}
\label{eq:Sss}
{\mathcal S}({\pmb P},{\pmb r})
=\frac{\int {\rm d}^4x_a \, {\rm d}^4x_b \, s_a \left(m_a{\pmb P}/(m_a+m_b),x_a\right) \,
  s_b\left(m_b{\pmb P}/(m_a+m_b),x_b\right) \, \delta({\pmb r}-{\pmb r}_a-{\pmb r}_b)}
{\int {\rm d}^4x_a \, {\rm d}^4x_b \, s_a\left(m_a{\pmb P}/(m_a+m_b),x_a\right) \, s_b\left(m_b{\pmb P}
/(m_a+m_b),x_b\right)} \, ,
\end{equation}
where $s_a({\pmb p},x)$ is the density for emitting particles of type $a$ with momentum
${\pmb p}$ from space-time point $x$. The spatial arguments are for the frame
moving at the center-of-mass velocity ${\pmb V}$.  According to Eq.\ (\ref{eq:corrmaster}), the
correlation data access the probability distribution in the relative
rest-frame separation only.  Information on emission duration can only be
inferred from the influence of that duration on the relative spatial
distribution \cite{Pratt:1986cc, Bertsch:1989vn, Csorgo:1989kq}.  Similarly,
single-particle distributions $s$ can be deduced only indirectly from data
(\ref{eq:corrmaster}) because of the folding in (\ref{eq:Sss}) with an
integration over the average particle position.  For a set of investigated
particles, the constraining of single-particle distributions may be enhanced by
an analysis of correlations (\ref{eq:corrmaster}) of all possible particle
pairs.

In the following, we concentrate on what might be directly determined from data, i.e.,
the determination of ${\mathcal S}({\pmb P},{\pmb r})$ given
measurement of ${\mathcal R}({\pmb P},{\pmb q})$.  Since each individual value
of ${\pmb P}$ nominally represents a different problem, we suppress the ${\pmb
  P}$ arguments in the functions.  As access to ${\mathcal S}$ is conditioned
on any deviations of $|\phi|^2$ from unity, we subtract unity from both sides of
(\ref{eq:corrmaster}) arriving at
\begin{equation}
\label{eq:RS3D}
{\mathcal R}({\pmb q})= \int {\rm d}^3r \, \left[|\phi^{(-)}({\pmb q},{\pmb r})|^2-1\right] {\mathcal S}({\pmb r}) \equiv \int {\rm d}^3r \, {\mathcal K} ({\pmb q},{\pmb r}) \,
{\mathcal S}({\pmb r})
 \, ,
\end{equation}
where we have utilized the normalization of ${\mathcal S}({\pmb P},{\pmb r})$.
Here, both ${\pmb q}$ and ${\pmb r}$ are evaluated in a frame
where ${\pmb P}=0$. It is seen that ${\mathcal R}$ is related to ${\mathcal S}$
through an integral transform where $|\phi^2({\pmb q},{\pmb r})|^2-1$ plays the
role of the kernel ${\mathcal K}$ of transformation.  The determination of
${\mathcal S}({\pmb r})$ from ${\mathcal R}({\pmb q})$ amounts then to the
inversion of the integral transform, often referred to as imaging
\cite{Verde:2003cx,Verde:2001md,Panitkin:2001qb,Brown:2000aj,
  Brown:1999ka,Brown:1997ku}.

Imaging has already been extensively performed for angle-averaged correlation
functions, ${\mathcal R}(q)$, which can provide angle-averaged distributions
${\mathcal S}(r)$. It is the goal of this paper to describe how one might
extract the angular information in ${\mathcal S}({\pmb r})$ by decomposing
${\mathcal R}({\pmb q})$ in terms of surface-spherical tesseral and cartesian harmonics.
A~parallel effort, employing tesseral harmonics in the case of pure
interference, is described in \cite{Brown:2005ze}.

The wavefunction squared $|\phi({\pmb q},{\pmb r})|^2$ depends, after summation
over spins, only on the magnitudes of $q$ and $r$ and on the angle
$\theta_{\pmb q \, r}$ between ${\pmb q}$ and ${\pmb r}$.  Thus, the square is
invariant under rotations.  As a consequence, the coefficients of expansion in
terms of spherical harmonics, for the correlation and source functions, are
directly related to each other~\cite{Brown:2000aj},
\begin{equation}
\label{eq:ylmmaster}
{\mathcal R}_{\ell m}(q)= 4\pi \int {\rm d}r \, r^2 \,
{\mathcal K}_\ell(q,r) \, {\mathcal S}_{\ell m}(r) \, .
\end{equation}
Here, the expansion coefficients and expansion for the correlation function are
defined by the relations:
\begin{eqnarray}
\nonumber
\label{eq:Rylm}
{\mathcal R}({\pmb q})&=& {\sqrt{4\pi}} \sum_{\ell m}{\mathcal R}_{\ell m}^* (q) \,
Y_{\ell m}(\Omega) \, ,\\
{\mathcal R}_{\ell m}(q)&\equiv&
\frac{1}{\sqrt{4\pi}}  \int {\rm d} \Omega_q  \, Y_{\ell m}(\Omega) \, {\mathcal R}({\pmb q}) \, ,
\label{eq:ylmmaster_defs}
\end{eqnarray}
with analogous expressions for the source function. The partial kernel
${\mathcal K}_\ell$ encodes the information from the wavefunction,
\begin{equation}
\label{eq:kernell}
{\mathcal K}_\ell(q,r) \equiv \frac{1}{2} \int {\rm d} \cos{\theta_{qr}} \, \left[
|\phi^{(-)}(q,r,\cos{\theta_{qr}})|^2-1\right] \, P_\ell(\cos{\theta_{qr}}) \, .
\end{equation}
According to the above, an $(\ell,m)$ coefficient of the correlation function
is tied to the coefficient of the source function with the same $(\ell,m)$. Thus, the source
restoration can be reduced to a series of one-dimensional inversions of the
integral transformations for individual $(\ell,m)$, employing kernels for
specific $\ell$. This sequence of one-dimensional inversions is computationally simpler
than a single three-dimensional
inversion.  The angular resolution of data would determine how far the
restoration can proceed in $\ell$.

One problem with the above is the emergence of complex coefficients for the real-valued
correlation and source
functions, lacking direct geometric interpretation.  The last issue is
circumvented by introducing an alternate real-valued basis in the space of
spherical angles, surface-spherical cartesian harmonics ${\mathcal
  A}_{\vec{\ell}}(\Omega)$ \cite{Danielewicz:2005qh,applequist89,applequist}.

Cartesian harmonics ${\mathcal A}_{\vec{\ell}}$ are linear combinations of
$Y_{\ell m}$ with different $m$ but the same $\ell = \ell_x + \ell_y +\ell_z$.
In terms of the directional unit vector ${\pmb n}$ pointing in the $\Omega$
direction, the cartesian harmonics are explicitly given by \cite{applequist89}
\begin{eqnarray}
{\mathcal A}_{\vec{\ell}} (\Omega)
&=&
\sum_{\substack{\vec{m} \\ 0 \le m_i \le \ell_i/2 }}
\left(-\frac{1}{2}\right)^m \, \frac{(2\ell-2m-1)!!}{(2\ell-1)!!}  \, \frac{\ell_x!}{(\ell_x-2m_x)! \, m_x!}
\nonumber
\\
&&\times
\frac{\ell_y!}{(\ell_y-2m_y)! \, m_y!} \,
\frac{\ell_z!}{(\ell_z-2m_z)! \, m_z!} \,
n_x^{\ell_x-2m_x} \, n_y^{\ell_y-2m_y} \, n_z^{\ell_z-2m_z} \, ,
\label{eq:Anseries}
\end{eqnarray}
where $m=m_x+m_y+m_z$ and $(-1)!!=1$. The leading term of ${\mathcal
  A}_{\vec{\ell}}$ is $n_x^{\ell_x} \, n_y^{\ell_y} \, n_z^{\ell_z}$, while the
subsequent terms ensure that $r^\ell \, {\mathcal A}_{\vec{\ell}}$
satisfies the Laplace equation and, thus, is a combination of the $Y_{\ell m}$
of different $m$.

The expansion in terms of cartesian harmonics are
defined with the relations
\begin{eqnarray}
\label{eq:cartesianmaster_exp}
{\mathcal R}({\pmb q}) & = &
\sum_{\vec{\ell}} \frac{\ell!}{\ell_x! \, \ell_y! \, \ell_z!} \,
{\mathcal R}_{\vec{\ell}} \, {\mathcal A}_{\vec{\ell}}(\Omega_{\pmb q})
=
\sum_{\vec{\ell}} \frac{\ell!}{\ell_x! \, \ell_y! \, \ell_z!} \,
{\mathcal R}_{\vec{\ell}} \, \hat{q}_x^{\ell_x} \, \hat{q}_y^{\ell_y} \, \hat{q}_z^{\ell_z}
\, ,
\\
\label{eq:cartesianmaster_coef}
{\mathcal R}_{\vec{\ell}}(q)&\equiv&
\frac{(2\ell+1)!!}{\ell!}\int \frac{d\Omega_q}{4\pi} \,
{\mathcal A}_{\vec{\ell}}(\Omega_q) \, {\mathcal R}({\pmb q})  \, .
\end{eqnarray}
In many situations the explicit form of ${\mathcal A}_{\vec{\ell}}$ is
not needed.  Just the existence of the cartesian harmonics and their properties
justify representing a function of the spherical angle in the second series in
(\ref{eq:cartesianmaster_exp}).

Since ${\mathcal A}_{\vec{\ell}}$ is a combination of $Y_{\ell m}$ with $\ell =
\ell_x + \ell_y +\ell_z$, the coefficients of expansion in terms of cartesian
harmonics for the source and correlation function are related to each other in
the same way as the coefficients of expansion in terms of tesseral harmonics in
Eq.\ (\ref{eq:ylmmaster}),
\begin{equation}
\label{eq:cartesianmaster}
{\mathcal R}_{\vec{\ell}}(q)= 4\pi \int {\rm d}r \, r^2 \,
{\mathcal K}_\ell(q,r) \, {\mathcal S}_{\vec{\ell}}(r) \, .
\end{equation}

In the next section, the structure of the kernel ${\mathcal K}_\ell$ in Eqs.\
(\ref{eq:ylmmaster}) and (\ref{eq:cartesianmaster}) is investigated for the
cases of pure identical-particle interference and for strong and Coulomb
interactions within a pair.  We find that all three classes of final-state
effects provide significant resolving power for different $\ell$.  Section
\ref{sec:cartesian} provides a detailed discussion of the properties of
cartesian harmonics.  The subsequent section presents a discussion of the
utility of higher $\ell$-moments and their relation to specific geometric
features of the source.  In particular, we show that the $\ell=1$ moments can
reveal an offset between the probability clouds for two different species, that
the $\ell=2$ moments can reveal the magnitude and orientation of axes in the
ellipsoidal approximation for a source, while the $\ell=3$ moments can reveal
the ``boomerang'' nature of a source.  In Sec.\ \ref{sec:examples}, the cases
of gaussian and blast wave sources are studied, as examples, in quantitative
detail.  In the final section, we summarize our results and discuss the
prospects and challenges expected in analyses employing the harmonics.

\section{Correlation Kernels ${\mathcal K}_\ell$}
\label{sec:kernels}

The ability to extract source ${\mathcal S}$ anisotropies depends on the
kernels ${\mathcal K}_\ell(q,r)$ defined in Eq.\ (\ref{eq:kernell}).  Data on
${\mathcal R}$ are obtained at a certain resolution in $q$ and suffer from
errors.  The theoretical relation (\ref{eq:corrmaster}) involves
approximations~\cite{Anchishkin:1997tb,Bertsch:1993nx}, in assuming that the
relative momentum $q$ is small compared to the scales characterizing the rest
of the system and in ignoring final-state effects other than those between
the two studied particles moving slow relative to each other.  If the kernels
${\mathcal K}_\ell$ drop rapidly as $\ell$ increases, for typical $q$-values
and for values of $r$ characteristic to the source function, there
will be no chance to determine details of the shape of ${\mathcal S}({\pmb r})$.
In the following we
analyze the situation for different types of final-state interactions at
different $\ell$, $q$ and $r$.

In the process of source restoration, the values of ${\mathcal
  R}_{\vec{\ell}}$ would be known at discrete values of $q$.  In parallel, the
source ${\mathcal S}_{\vec{\ell}}(r)$ might be discretized in $r$ or decomposed in some basis.  The source restoration would then amount to the inversion of a matrix
out of ${\mathcal K}_\ell(q,r)$, cf.\ \cite{Brown:1997sn,Brown:2000aj}.
An inability to extract source features might be signalled by the proximity
of matrix eigenvalues to zero, resulting in an
instability of the inversion.  Such a situation could be
encountered when extracting details of ${\mathcal S}_{\vec{\ell}}$ in an
$r$-region where ${\mathcal K}_\ell$ lacks resolving power.  Practical
experience in matrix-inversion strategy, in analyzing data, has been gained for $\ell=0$
\cite{Verde:2003cx,Verde:2001md,Panitkin:2001qb,Brown:2000aj,Brown:1999ka,Brown:1997ku}.
Importantly, this strategy has revealed non-Gaussian features in ${\mathcal
  S}_{\ell=0}(r)$, which were especially significant for proton sources in intermediate
energy collisions.  More common in the literature than the source
discretization, has been a parameterization of the source which is then fit to ${\mathcal R}$.
When parameterizing the
source in three-dimensions, the benefit regarding angular moments can be in
understanding a~systematic of the moments, ${\mathcal R}_{\vec{\ell}}$ and
${\mathcal S}_{\vec{\ell}}$, as a function of $\vec{\ell}$.  With regard to the
fit strategy, the features of ${\mathcal K}_\ell(q,r)$ determine what
parameters of ${\mathcal S}$ may be potentially constrained by data.

In the first subsection we discuss kernels for the case of pure identity interference between
spin-0 bosons.  The two subsequent subsections are dedicated to the Coulomb and the
strong interactions. Before launching into a detailed discussion of specific classes of
interactions, we emphasize a general observation that a three-dimensional source
$S_{\vec{\ell}}(r)$, expanded in a Taylor series around ${\pmb r}=0$, behaves as $r^\ell$ for small $r$.
Consequently, there is not much to be discerned about the source at small $r$ for $\ell \ge 1$,
particularly for large $\ell$, and the structure of $K_{\ell \ge 1}$ at small $r$ will not be overly
important. On the other hand, much could be learned about the angular structure of the source at large $r$.
However, when the large-$r$ region gets mapped onto small $q$, the ability to
gain angular information becomes limited by the experimental resolution capabilities.
Thus, in practice, the most important features of ${\mathcal K}_{\ell \ge 1} (q,r)$ are
those at moderate $q$ combined with moderate to high $r$. The naive expectation regarding
strong interactions may be that when an $s$-wave dominates the interaction, there is no
possibility to discern source shapes. We shall see that this is not the case.

\subsection{Kernels for Spin-0 Boson Identity-Interference}
\label{subsec:kernel_bosons}

For non-interacting identical spin-0 bosons, the squared relative wave function
is
\begin{equation}
\label{eq:phi2idb}
|\phi(q,r,\cos{\theta_{qr}})|^2=1+\cos{(2qr\cos{\theta_{qr}})} \, ,
\end{equation}
where the relative momentum convention is that ${\pmb q}$ stands for the
momentum of one of the particles in the center-of-mass frame.  In this case,
the three-dimensional transformation (\ref{eq:RS3D}) becomes the cosine Fourier
transform \cite{Brown:1997ku}.  Since the cosine term in (\ref{eq:phi2idb}) may
be represented as the real part of ${\rm e}^{2 {\rm i} {\pmb q}\cdot{\pmb r}}$
and the latter may be expanded in Legendre polynomials and spherical Bessel
functions, the ${\mathcal K}_\ell$-transformations (\ref{eq:ylmmaster}) and
(\ref{eq:cartesianmaster}), with (\ref{eq:kernell}), emerge as spherical-Bessel
transforms (see also \cite{Brown:2005ze}):
\begin{equation}
\label{eq:Klinterf}
{\mathcal K}_\ell(q,r)=\left\{
\begin{array}{cl}
(-1)^{\ell/2} j_\ell(2qr) \, ,&\mbox{for even $\ell$} \, ,\\
0 \, ,&\mbox{for odd $\ell$} \, . \end{array} \right.
\end{equation}

The properties of $j_\ell(2qr)$, as a function of $qr$, are as follows.  At low
$qr$, the function starts out in the power form, $j_\ell(2qr) \approx
(2qr)^\ell /\ell!!$, which is followed, for $\ell \ge 2$, by a maximum at $qr
\sim \frac{\ell}{2} + \frac{1}{4}$ and by a subsequent alternation at the
approximate period of $\pi$. This has the following consequences for the
ability to discern the source angular components $S_\ell$.  If $R$ is a scale
characterizing $S_\ell$, $S_\ell$ will have little impact on
${\mathcal R}_\ell$ if, at $\ell \ge 2$, $qR$ falls into the hole in $j_\ell$
at low argument values.  Likewise, $S_\ell$ will have little impact on
${\mathcal R}_\ell$, if many oscillations of $j_\ell$ are averaged over in the
spatial integration over the source.  The maximal impact is expected if $qR$
falls into the region of the first maximum in $j_\ell$ and, thus, most
information on $S_\ell$ may be expected in ${\mathcal R}_\ell$ at $q \sim (\ell
+ \frac{1}{2})/(2R)$.  Otherwise, the coefficients ${\mathcal R}_\ell$ are
expected to behave as $q^\ell$ for low $q$ and vanish at large~$q$.

As characteristic for Fourier transforms, $q$ and $r$ enter the kernels as a
product, and the low-$r$ region of the source gets predominantly mapped onto the
high-$q$ region of the correlation function.  Conversely, the high-$r$ region of
the source function mainly affects the low-$q$ region of the correlation function.
The inverse to the cosine Fourier transform is again the cosine Fourier transform
and the inverse to the spherical Bessel transform is the spherical Bessel transform.
Given the Bessel function completeness relation,
\begin{equation}
\int_0^\infty {\rm d}q \, q^2 \, j_\ell(2qr) \, j_\ell(2qr') =\frac{\pi}{16} \, \delta(r-r') \, ,
\end{equation}
we can introduce an inverse kernel
\begin{equation}
{\mathcal K}_\ell^{-1}(r,q)=\left\{
\begin{array}{cl}
\frac{(-1)^{\ell/2}}{\pi^3} j_\ell(2qr) \, ,&\mbox{for even $\ell$} \, ,\\
0 \, ,&\mbox{for odd $\ell$} \, , \end{array} \right.
\end{equation}
and obtain
\begin{equation}
{\mathcal S}_{\vec{\ell}}(r)= 4\pi \int {\rm d}q \, q^2 \,
{\mathcal K}_\ell^{-1}(r,q) \, {\mathcal R}_{\vec{\ell}}(q) \, .
\end{equation}

\subsection{Kernels for Repulsive Coulomb Interaction}
\label{sec:kernel_coul}

The Coulomb final-state interaction also provides means for determining
${\mathcal S}({\pmb r})$ from ${\mathcal R}({\pmb q})$ \cite{Pratt:2003ar,kim}.
In the Coulomb case, there is no straightforward way to invert the relation to
the source function as in the case of pure interference.  To understand how
${\mathcal S}$ impacts ${\mathcal R}$ for the Coulomb interaction, the
structure of the kernel must be considered in detail.

In the classical limit of a repulsive Coulomb interaction, the kernel depends
solely on $\theta_{qr}$ and on the ratio of the pair relative-energy to the pair-Coulomb-energy at emission:
\begin{equation}
\label{eq:xdef}
x= \frac{Z_1 \, Z_2 \, e^2}{ r} \, \frac{2 \mu}{q^2} = \frac{r_C}{r} \, ,
\end{equation}
where $\mu$ is the reduced mass and $r_C$ is the radius of the
Coulomb barrier in the pair interaction.  Quantal effects become
important for $qr/\hbar$ of the order of unity.  In the quantal
situation, the kernel also depends on the dimensionless variable
$qr/\hbar$. The Gamow-factor parameter is then given by the
parameter product $\eta = x \, qr/2 = Z_1 \, Z_2 \, e^2 / v$ and a
further parameter multiplication yields $x \, (qr)^2 = 2r/a_0$,
where $a_0 = 1/Z_1 \, Z_2 \, e^2 \, \mu$ is the Bohr radius.

In the classical limit, the squared wavefunction
represents the ratio of initial and final spatial densities
during motion to the detectors for particles of a given initial momentum,
or, due to the conservation of phase-space density, the inverse ratio of
the momentum densities at a fixed initial separation:
\begin{equation}
\label{eq:coulapprox} |\phi^{(-)}(q,r,\cos{\theta_{qr}})|^2
\rightarrow \frac{{\rm d}^3 \, r_\infty }{{\rm d}^3\, r}  =
\frac{{\rm d}^3 \, q_0}{{\rm d}^3 \, q}
=\frac{1+\cos\theta_{qr}-x} {\sqrt{(1+\cos{\theta_{qr}}-x)^2-x^2}}
\, \Theta(1+\cos{\theta_{qr}}-2x) \, .
\end{equation}
Here, ${\pmb q}_0$ represents the relative momentum of the particles at
separation ${\pmb r}$, while ${\pmb r}_\infty$ represents a remote separation of
the particles at some instant when the relative momentum has approached ${\pmb q}$.
The r.h.s.\ result in (\ref{eq:coulapprox}) follows from Coulomb-trajectory
considerations \cite{Pratt:2003ar}.  The implications of this result simplify
\cite{Danielewicz:2005qh} for $\ell=0$ and in the high-energy ($x \ll 1$) and
low-energy ($x \gg 1$) limits.  Thus, for $\ell=0$, the partial kernel is
\begin{equation}\label{eq:kernel_Coul_0}
 {\mathcal K}_0 (q,r) = \frac{q_0^2}{q^2} \frac{\text{d}q_0}{\text{d}q} - 1
 = \frac{q_0}{q} - 1 = \frac{\Theta(x)}{\sqrt{1-x}} - 1 \, .
\end{equation}

\begin{figure}
\centerline{\includegraphics[width=0.4\textwidth]{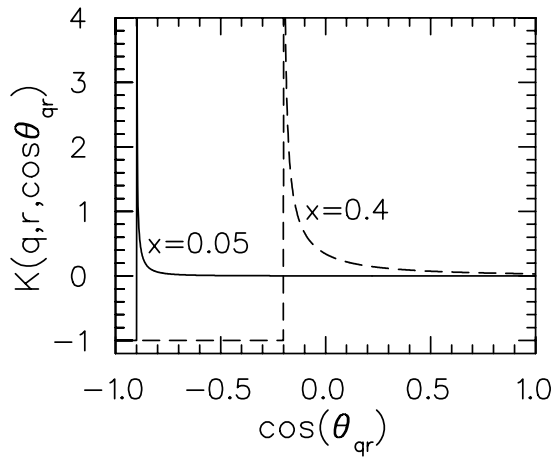}}
\caption{\label{fig:classcoulkernel} Kernel, ${\mathcal
    K}(q,r,\cos\theta_{qr})= |\phi^{(-)}|^2-1$, for classical repulsive Coulomb
  interaction, shown as a function of $\cos{\theta_{qr}}$ at a couple of values
  of the dimensionless parameter, $x=r_C/ r$, where $r_C=2 \mu Z_1 Z_2 e^2 /
  q^2$.  The structure around $\cos{\theta_{qr}}=-1$ is due to the deflection
  of classical trajectories directed towards the center of the repulsive
  Coulomb interaction.  As the relative momentum within the pair increases, at
  fixed $r$, the range of backward angles where the kernel is significant
  shrinks.}
\end{figure}
The three-dimensional kernel ${\mathcal K}$ is illustrated in Fig.\
\ref{fig:classcoulkernel}, as a function of $\cos{\theta_{qr}}$, for two
$x$ values. In the high energy limit of $x \ll 1$, the wavefunction squared
$|\phi|^2$ is close to unity for most values of $\cos{\theta_{qr}}$.  In that
limit, the trajectories emerging from ${\pmb r}$ are practically straight lines.
The exception are the trajectories aiming towards the center of interaction,
that cannot penetrate the region of radius $r_C$.  The trajectory directions
with $\cos{\theta_{qr}} < -1 + 2x$ get shadowed.  (The factor of~2 in front of
$x$ stems from the fact that trajectories get deflected both during the motion
towards and away from the barrier.)  The deflected trajectories pile up
primarily outside of the shadowed region.  The pile-up produces an integrable
singularity in the kernel as shown by the peaks in Fig.\
\ref{fig:classcoulkernel}.  At low $\ell$, the Legendre polynomials have a
limited angular resolution.  Given that limited resolution, for the purpose of
the angular integration in (\ref{eq:kernell}), the angular structure in
$|\phi|^2$, in vicinity of the backward direction $\cos{\theta_{qr}}=-1$, may
be approximated by a $\delta$-function multiplied by the integral over the
structure:
\begin{equation}
  \label{eq:Coulwfasy}
  |\phi^{(-)}(q,r,\cos{ \theta_{qr}})|^2 \approx 1
-\frac{x}{2}\, \delta(1+\cos{\theta_{qr}}) \, , \hspace*{1em}
\mbox{for $x \cdot \max{(1,\ell)} \ll 1$} \, .
\end{equation}
Since $P_\ell(-1)=(-1)^\ell$, the high-energy kernel can be then approximated as
\begin{equation}
\label{eq:kl_coul_tail}
{\mathcal K}_\ell(q,r) \approx (-1)^{\ell+1} \, x/2 \, , \hspace*{1em}
\mbox{for $x \ll 1/\max{(1,\ell)}$ and $qr \gg \max{(1,\ell)}$} \, .
\end{equation}

For more implications of the classical limit (\ref{eq:coulapprox})
for kernels ${\mathcal K}_\ell$, see
ref.~\cite{Danielewicz:2005qh}.  The classical limit of the
Coulomb interaction is of interest in the context of emission of
intermediate-mass fragments~(IMF) from low-energy central nuclear
reactions \cite{kim}.  The practical condition of applicability
for the classical approach is that $QR \gg \max{(1,\ell)}$, where
$Q$ is the characteristic momentum which yields $r_C \sim R$. With
$Z \sim A/2$, this produces, at low $\ell$, the condition $(e^2 \,
A^3 \, m_N \, R)^{3/2} \gg 2$.  Given a typical value of $R \sim 5
\, \mbox{fm}$ for heavy ion reactions, this amounts to $A^{3/2}
\gg 5$ and the classical limit is met at the fragment mass of $A
\gtrsim 6$.  IMF correlations are commonly expressed in terms of
reduced velocity $v_{\rm
  red}=q/\mu (Z_1+Z_2)^{1/2}$.  With $Z \sim A/2$, $r_C \sim 2 e^2/m_N v_{\rm
  red}$ and $R \sim 5 \, \mbox{fm}$, the high-energy limit of $x \ll 1$ for the
kernel is reached at $v_{\rm red} \gtrsim 0.04 \, c$.

At small $qr$ in an interacting system, quantum effects become
important as phase-space delocalization produces diffraction and
tunneling. The diffraction affects the angle of an emerging
particle at the level of $1/qr$, smearing out the angular
structure in the kernel.  Given the resolution of Legendre
polynomials changing with $\ell$, the diffraction effects become
consequently important for $qr \lesssim \max{(\ell,1)}$, with the
kernel ${\mathcal K}_\ell$ then being reduced on average.  The
tunneling into the region of $x > 1$ becomes significant down to
$r=0$, for $2 \eta = x qr \equiv q \, r_C \lesssim 1$.  For singly
charged particles, such as identical pions, this corresponds to $v
\gtrsim 0.02 \, c$.  Unlike the classical
expression~(\ref{eq:coulapprox}), the Coulomb wavefunction squared
\begin{equation}\label{eq:Coulombsq}
  |\phi^{(-)} (q,r, \cos{ \theta_{qr}})|^2 = G(\eta) \,
  |M({\rm i} \eta , 1 , -{\rm i}qr (1 + \cos { \theta_{qr}})|^2
\end{equation}
is analytic around $r=0$ in the argument $r (1 + \cos { \theta_{qr}}) \equiv r
+ {\pmb r}\hat{\pmb q}$.  In the quantal expression,
\begin{equation}\label{eq:Gamow}
  G(\eta) = \frac{2 \pi \eta}{{\rm e}^{2 \pi \eta}-1}
\end{equation}
is the Gamow factor. The terms in the Taylor series for the squared
wavefunction (\ref{eq:Coulombsq}) behave as $r^\ell \, (1 +
\cos{\theta_{qr}})^\ell$.  Since the highest power of $\cos{\theta_{qr}}$ for
an $\ell$'th term is $\ell$, the quantal Coulomb kernel ${\mathcal K}_\ell$
from~(\ref{eq:kernell}) behaves around $r =0$ as $r^\ell$, similarly to the identity-interference kernel (\ref{eq:Klinterf}).  The power behavior, associated with the quantal
delocalization, holds up to $r \lesssim \min{(a_0,q^{-1})}$. The behavior
extends past the classical Coulomb barrier when $r_C \lesssim
\min{(a_0,q^{-1})}$, leading to a further dampening of ${\mathcal K}_\ell$ at
$\ell \ge 1$ in the barrier region.

\begin{figure}
\centerline{\includegraphics[width=0.7\textwidth]{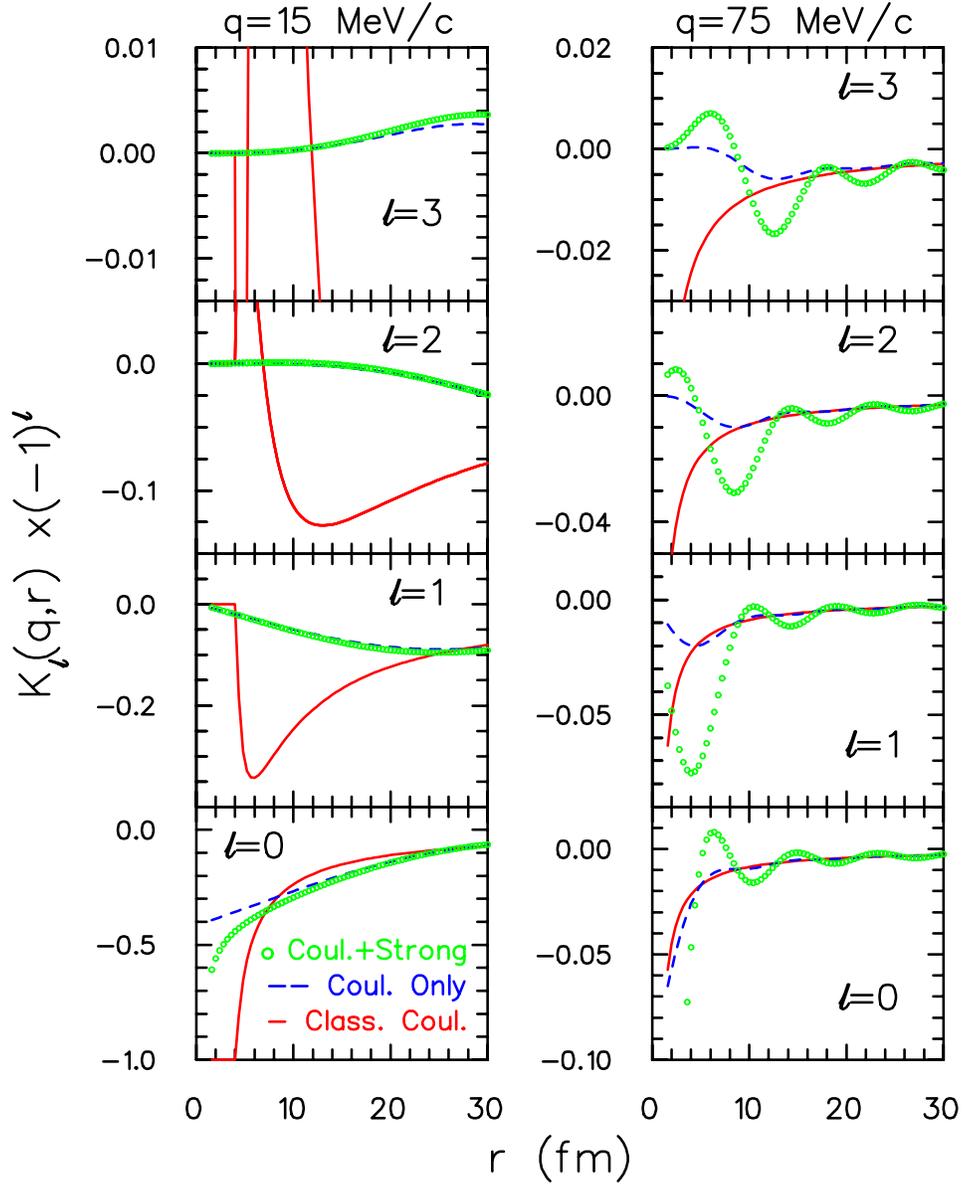}}
\caption{\label{fig:kernel_pkplus} (Color online) Kernels ${\mathcal K}_\ell$, multiplied by
  $(-1)^\ell$, for $pK^+$ interactions at $q=15 \, \mbox{MeV/c}$ (left panels)
  and $q=75 \, \mbox{MeV/c}$ (right panels), shown as a function of $1.5 \,
  \mbox{fm} < r < 30 \, \mbox{fm}$, for $\ell \le 3$.  Classical and quantal
  Coulomb kernels and quantal kernels with inclusion of strong interactions are
  represented by solid lines, dashed lines and symbols, respectively.  The
  vertical scale may change from a panel to panel to emphasize details.
  Kernels are suppressed at the lowest $r$ for the sake of clarity.}
\end{figure}

As an example, in Fig.~\ref{fig:kernel_pkplus}, we consider
${\mathcal K}_\ell$ kernels, for the $pK^+$ interaction.  The
kernels are shown for $\ell = (0-3)$, at relative momenta of $q=15$ (left
panels) and 75 MeV/c (right panels), as a function of distance~$r$.  Regions of
higher and lower relative momenta within a correlation function
$\mathcal{R}(q)$ are, typically, used to determine, respectively, about the short
and long-rage features of the source $\mathcal{S}(r)$.  The solid lines, dashed
lines and symbols, represent the classical Coulomb kernels, quantal Coulomb
kernels and quantal kernels with inclusion of strong interactions.

We first address the Coulomb kernels at 75~MeV/c.  At large $r$ in
the right panels of Fig.~\ref{fig:kernel_pkplus}, the classical
Coulomb kernels ${\mathcal K}_\ell$ are, up to a sign, independent
of $\ell$ and fall off as $1/r$, consistently with
Eq.~(\ref{eq:kl_coul_tail}). Differences emerge at low $r$.  At
$\ell \ge 1$ the different classical kernels ${\mathcal K}_\ell$
switch sign $(\ell - 1)$, number of times, cf.\
\cite{Danielewicz:2005qh}.  With the distance scale of $r_C$
being, though, a fraction of a Fermi, $r_C = 0.17 \, \mbox{fm}$ at
$q=75 \, \mbox{MeV/c}$, the structures associated with the sign
switching are all pushed to very small $r$.  Those sign-switching
structures cannot be seen in the figure as the kernels at $r < 1.5
\, \mbox{fm}$ are suppressed for the sake of clarity.  For the
quantum kernels, given $q^{-1} = 2.6 \, \mbox{fm}$, those kernels
are seen to fluctuate with $r$ around the classical kernels at $r
\gg \max(1,\ell) \, q^{-1}$.  At lower $r$ and $\ell \ge 1$
quantal effects dampen the kernels, and in particular, at very low
$r$ the classical sign-switching in the kernels gets completely
erased with the kernels behaving as~$r^\ell$.  With inclusion of
the effects of strong interactions in the kernels, additional
oscillations are observed with $r$ within the figure, as the
asymptotic behavior of the partial wavefunctions changes for low
angular momenta, producing wave-function distortions at moderate
distances.  We will focus on strong interactions in the next
subsection.

At $q=15 \, \mbox{MeV/c}$, the classical Coulomb radius rises to $r_C= 4.1 \,
\mbox{fm}$, on one hand pushing the asymptotic high-energy region out of the
$r$-range shown in Fig.~\ref{fig:kernel_pkplus}, and, on the other hand, making
some of the low-$r$ details of the classical Coulomb kernels visible within the shown
range.  Also, since $q^{-1}=13.1 \, \mbox{fm}$ then, much of the region for the left
panels of Fig.~\ref{fig:kernel_pkplus} is dominated by quantal effects.  The
kernels with inclusion of strong interactions get to be very close to the
Coulomb kernels, since the phase shifts, characterizing changes in the
partial-wave asymptotic behavior, are close to zero at low~$q$.

The quantal suppression of the kernels at $qr \lesssim \ell$, progressing with
$\ell$, makes it difficult to learn about the directional characteristics of a
source from the particle correlations at low $q \lesssim R^{-1} = 40 \,
\mbox{MeV/c}$ for $R=5 \, \mbox{fm}$.  Small kernels produce small
contributions to the correlation which are difficult to detect experimentally.
At the other end, at high $qr$, while the kernels become independent of $\ell$,
they also become small, falling off as $1/(q^2 \, r)$, cf.\
(\ref{eq:kl_coul_tail}).  As a consequence, an optimal bracket of $q$ for
accessing the directional characteristic of sources with typical size emerges,
such as $q \sim (30-100) \, \mbox{MeV/c}$ for the $pK^+$ system.

We have seen the importance of the analyticity of $|\phi|^2$ in $r$ around
$r=0$, which implied the $r^\ell$ suppression of the ${\mathcal K}_\ell$ ($\ell
\ge 1$) kernels at low~$r$. The analyticity in $q$ is of interest in the
context of the analysis of correlation functions as a function of~${\pmb q}$.
In the case of pure identity interference, the wavefunction squared
(\ref{eq:phi2idb}) is analytic in~${\pmb q}$ which makes the kernels
from (\ref{eq:RS3D}) analytic in~${\pmb q}$.  The resulting correlation
coefficients ${\mathcal
  R}_\ell$ are also analytic and behave as $q^\ell$ at small~$q$.  In
  contrast, because of the general dependence on the parameter
\begin{equation}\label{eq:eta}
  \eta = \frac{Z_1 \, Z_2 \, e^2 \, \mu}{q} \, ,
\end{equation}
inversely proportional to~$q$, and because of the appearance of
the Gamow factor (\ref{eq:Gamow}), the square of
the Coulomb wavefunction (\ref{eq:Coulombsq}) lacks analyticity in
${\pmb q}$ around $q=0$.  Still, it may be tempting to factor out
the Gamow factor, as has been done in the analysis of pion
correlations, arriving, with $|\phi|^2/G \equiv |M|^2$, at a
Gamow-corrected kernel.  The corrected kernel remains finite at a
given $r$ and $\theta_{rq}$, when $q \rightarrow 0$, but retains a
dependence in this limit on $\cos{\theta_{qr}}$ for any finite
value of~$r$. Thus, $|M|$ reaches a different value depending on
the side from which $q=0$ is reached and lacks therefore
analyticity in~${\pmb q}$.  Correspondingly, even the
Gamow-corrected correlation functions from~(\ref{eq:RS3D}) lack
analyticity in~${\pmb q}$.  The Gamow-corrected kernels ${\mathcal
K}_\ell$ tend to finite values as $q \rightarrow 0$, rather than
behaving as~$q^\ell$.

As to the high-$q$ limit of the Coulomb kernels, while the resolving power
becomes independent of $\ell$, the accessed information on the source
represents one moment only per correlation coefficient, as
\begin{eqnarray}
{\mathcal R}_{\ell m}(q)
& \underset{ q \rightarrow \infty}{\approx} &
\frac{2 \sqrt{\pi} (-1)^\ell \, Z_1 \, Z_2 \, e^2 \, \mu }{q^2}
\left\langle \frac{1}{r} \, Y_{\ell m}^*(\Omega)
\right\rangle \, ,\\
{\mathcal R}_{\vec{\ell}}(q)
& \underset{ q \rightarrow \infty}{\approx} & \frac{(-1)^\ell \, Z_1 \, Z_2 \, e^2 \, \mu }{q^2} \left\langle
\frac{1}{r} \, {\mathcal A}_{\vec{\ell}}(\Omega)
\right\rangle \, ,
\end{eqnarray}
where
\begin{equation}
\langle F \rangle=\int {\rm d}^3 r \, {\cal S}({\pmb r}) \, F({\pmb r}) \, .
\end{equation}
Since the classical approximation is accurate at large $qr$, this
relation is unchanged by quantum considerations.  Working within
the classical approximation, one might be tempted to continue the
$r_C/r$ kernel expansion, producing expansion of the correlation
in powers of $1/q^2$.  The next expansion term could give access
to the moment of the source function with a factor of $1/r^2$ in
place of $1/r$.  However, beyond the leading term, the expansion
gets altered by quantum effects making the expectation invalid for
lighter particle pairs such as $pK^+$.

\subsection{Kernels for Short-Range Interactions}
\label{subsec:kernel_strong}

Strong interactions between two particles also provide leverage for extracting
source information from measured correlations and are particularly effective
when the strong-interaction cross-sections are large \cite{Pratt:2003ar}.  In
analysis of the effects of the interaction, the two-particle scattering
wavefunction may be conveniently decomposed into angular-momentum eigenstates,
here with any intrinsic spins suppressed:
\begin{equation}\label{eq:wfout}
  \phi^{(-)}(q,r, \cos{\theta_{qr}}) =
  \sum_{\ell_1} a_{\ell_1} \,  \frac{R_{\ell_1}(r)}{r} \, P_{\ell_1}(\cos{\theta_{qr}})
  \underset{r \rightarrow \infty}{\approx}
  {\rm e}^{{\rm i}qr \cos{\theta_{qr}}} + f^*(\pi - \theta_{qr}) \, \frac{{\rm e}^{-{\rm i}qr}}{r} \, .
\end{equation}
The coefficients $a_{\ell}$ in front of the radial wavefunctions $R_{\ell}$ are
adjusted to make the large-$r$ {\em outgoing} wave-contributions to
$\phi^{(-)}$ identical to those for a wave in the absence of short-range
interactions.  The explicit asymptotic form of the wavefunction, on the r.h.s.,
is shown for the case with no Coulomb interactions.  The amplitude $f \equiv
f^{(+)}$ is the scattering amplitude associated with the more standard
scattering wavefunction $\phi^{(+)}$ where conditions are imposed onto {\em
  incoming} wave-contributions in the asymptotic region.  The amplitude factors
in the two wavefunctions are related with $f^{(-)} (\theta)= \left[ f^{(+)}
  (\pi - \theta) \right]^*$.

The three-dimensional kernel from (\ref{eq:wfout}) is
\begin{equation}
\label{eq:kernelstrong}
{\mathcal K}(q,r, \cos{\theta_{qr}}) \equiv  |\phi^{(-)}|^2-1 = \frac{1}{r^2}
\sum_{\ell_1 \, \ell_2} a_{\ell_1} \, a_{\ell_2}^* \, R_{\ell_1}(r) \, R_{\ell_2}(r) \,
P_{\ell_1}(\cos{\theta_{qr}}) \, P_{\ell_2}(\cos{\theta_{qr}}) -1
\end{equation}
Since most shape information is contained at intermediate to large
$r$, the large $r$ limit is especially insightful and can be found
from Eq.~(\ref{eq:wfout}),
\begin{eqnarray}
  \nonumber
  {\mathcal K}(q,r , \cos{\theta_{qr}})&\underset{r
  \rightarrow \infty}{\approx}& \frac{1}{r^2} \, \frac{{\rm d}
  \sigma}{{\rm d} \Omega} (\pi - \theta_{qr})  -\frac{{\rm Im}\
  f(\pi-\theta_{qr})}{r} \, \sin[qr(1+\cos\theta_{qr})]
  \\ \label{eq:kernelR}
  &&  +\frac{{\rm
  Re}\ f(\pi-\theta_{qr})}{r} \, \cos[qr(1+\cos\theta_{qr})]
  \\
  \label{eq:kernelsr} & \rightarrow & \frac{1}{r^2} \, \frac{{\rm d}
  \sigma}{{\rm d} \Omega} (\pi - \theta_{qr}) - \frac{\sigma}{2 \pi
  r^2} \, \delta( 1 + \cos{\theta_{qr}}) \, .
\end{eqnarray}
In arriving at (\ref{eq:kernelR}) and (\ref{eq:kernelsr}), we used
an expression for the scattering cross section, $|f|^2=d \sigma/d
\Omega$, and the optical theorem, $\text{Im} \, f(0) = q \sigma /4
\pi $.  Further, in obtaining the expression (\ref{eq:kernelsr})
we have exploited the fact that for $qr \rightarrow\infty$, under
an ${\pmb r} $-integration, the respective products in
(\ref{eq:kernelR}) approach the limits: $qr\sin(qrx) \rightarrow
\delta(x)$ and $qr\cos(qrx) \rightarrow \ 0$.  The result
(\ref{eq:kernelsr}) has a simple geometric interpretation.  The
first term represents the probability that a particle aimed
directly at the scatterer reflects into the direction of~${\pmb
q}$, while the negative term represents the shadowing  of a
spherical angle, $\sigma/4 \pi r^2$, around the backward
direction.
\begin{figure}
\centerline{\includegraphics[width=0.45\textwidth]{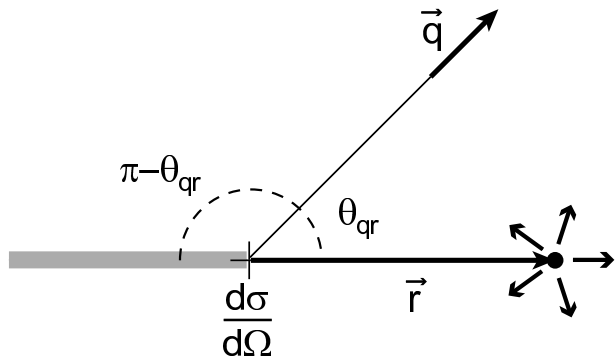}}
\caption{\label{fig:scattering} In the asymptotic region of $r \rightarrow
  \infty$, for short-range interactions, the kernel $K$ represents changes to
  the flux of particles emerging at relative momentum ${\pmb q}$, produced by scattering.
  Here, the particles were initially separated by ${\pmb r}$.}
\end{figure}

The partial kernels ${\mathcal K}_\ell$, respectively from
Eq.~(\ref{eq:kernelstrong}) and (\ref{eq:kernelsr}), are
\begin{eqnarray}
  \label{eq:kelsr}
  {\mathcal K}_\ell (q,r) &=& \frac{1}{r^2} \sum_{\ell_1 \, \ell_2}
  \begin{pmatrix}
  \ell & \ell_1 & \ell_2 \\
  0 & 0 & 0
  \end{pmatrix}^2 \, a_{\ell_1} \, a_{\ell_2}^* \, R_{\ell_1}(r) \, R_{\ell_2}(r)- \delta_{\ell 0} \\
  &&\underset{r \rightarrow \infty}{\approx} \frac{(-1)^\ell}{r^2}
  \, \left[ \frac{1}{2} \int \text{d}\cos{\theta} \, P_\ell
  (\cos{\theta}) \, \frac {\text{d} \sigma}{\text{d}\Omega} (\theta)
  - \frac{\sigma}{4 \pi} \right] \, ,
  \label{eq:klsr}
\end{eqnarray}
where we have made use of the Wigner $3j$-symbol.  In the
asymptotic region, all $\ell \ge 1 $ kernels decrease as $1/r^2$
and are proportional to the scattering cross section. Larger cross
sections, such as associated with resonances, will lead to larger
kernels at a given~$q$. Otherwise, the kernels are sensitive to
the angular dependence of the cross sections.  For $q$ in the
range for which correlations are typically studied, cross sections
are usually fairly isotropic. Thus, the first term in the brackets
on the r.h.s.\ of (\ref{eq:klsr}) decreases quickly as $\ell$
increases, yielding
\begin{equation}\label{eq:kllarge}
  {\mathcal K}_\ell (q,r) \approx \frac{(-1)^{\ell+1} \, \sigma}{4 \pi r^2} \, ,
  \hspace*{1.em} \text{for large $r$ and $\ell$.}
\end{equation}
Apart from the sign, this kernel is independent of $\ell$, similarly to the
Coulomb kernel (\ref{eq:kl_coul_tail}).  In analogy to the classical Coulomb
results for individual~$\ell$, the asymptotic results (\ref{eq:klsr}) require
$qr \gg \ell$ for their validity.  An additional condition for validity is that
$r/d \gg \ell$ where $d$ is the interaction range.

Regarding the limiting short-range~(\ref{eq:kllarge}) and Coulomb
(\ref{eq:kl_coul_tail}) ${\mathcal K}_\ell$-results, their
independence from $\ell$ results from the kernels being
exclusively influenced by the shadowing of particles by each
other.  The effect of shadowing decreases with distance between
the particles, but more slowly for the Coulomb than for purely
short-range interaction.  In the case of short-range interactions,
the kernel falls off as $\sigma/r^2$, while for the Coulomb forces
the kernels fall off as $r_C/r$, where $r_C=2\mu Z_1Z_2e^2/q^2$ is
the radius of the Coulomb barrier. Since $r_C$ explicitly
decreases with $q$, Coulomb kernels decrease monotonically with
increasing $q$, whereas for the strong interaction, the $q$
dependence on the kernel tends to follow the energy dependence of
the cross section. Strong interactions can thus provide resolving
power at large $q$, if the cross sections are large, e.g.\ at
resonance energies.

When the two terms in the asymptotic form of the kernel
(\ref{eq:kernelsr}) are integrated over full spherical angle,
those terms exactly cancel, reflecting the fact that, for a
short-range interaction, the particles which are shadowed for
$\cos\theta_{qr}=-1$ reappear at other angles at the same
magnitude of relative momentum.  Consequently, the $\ell=0$
coefficient for the $1/r^2$ asymptotic fall-off in (\ref{eq:klsr})
vanishes, implying that the ${\mathcal K}_0$ kernel
from~(\ref{eq:kelsr}) vanishes faster than $1/r^2$ as $r
\rightarrow \infty$.  These results may be contrasted with what is
found for the Coulomb interactions. The large-$r$ kernel Coulomb
kernel from~(\ref{eq:Coulwfasy}) yields a finite result when
integrated over the spherical angle.  In direct consequence, the
$\ell=0$ kernel is finite in the asymptotic region,
Eqs.~(\ref{eq:kernel_Coul_0}) and (\ref{eq:kl_coul_tail}). The
difference, between the short-range and Coulomb cases, is due to
the fact that a Coulomb interaction does not just change the
orientation of the relative momentum but also makes the momentum
magnitude different from the asymptotic value at any finite
distance from the interaction center.

At low relative momenta~$q$, $s$-wave scattering is
likely to dominate.  An important result, visible already in the
asymptotic limit, Eqs.~(\ref{eq:klsr}) and (\ref{eq:kllarge}), is
that the $s$-wave scattering contributes to the kernels at
all~$\ell$.  At the level of Eq.~(\ref{eq:kelsr}), it is seen as
associated with the interference between the waves.  At resonance,
the radial wavefunctions $R_\ell$ will be particularly large in the
near zone giving rise to enhanced ${\mathcal K}_\ell$ there, including
$\ell=0$.
At low $q$ and finite $\ell$, the effects of the centrifugal barrier can be important.
When $q$ is so low that the barrier is encountered outside of the strong-interaction
range, $qr < \ell + \frac{1}{2}$ at $r >d$, the radial wavefunctions are approximately
given in terms of the spherical Bessel functions, $R_\ell/r \propto j_\ell (qr)
\propto (qr)^\ell$.  With $R_0$ approaching a constant value as a function of both $q$ and $r$ in
the region $d < r \ll 1/q$, at $q \rightarrow 0$, given the triangle inequality
for the $\ell$-values
in the $3j$ symbol in~(\ref{eq:klsr}), the strong-interaction kernels will behave as ${\mathcal K}_\ell(q,r)
\propto (q  r)^\ell (1 + b/r)$ for $\ell \ge 1$ in the region.
The kernels will be suppressed then at low $q$ and low $r$ as well for larger~$\ell$.
A short-range repulsion may enhance suppression at lowest $r < d$ beyond that produced by
the centrifugal barrier.

Outside of the interaction range, at $r > d$, the wavefunctions
$R_\ell$ are fully determined by the respective strong-interaction
phase-shifts, both in the absence and presence of Coulomb
interactions. We exploit this  \cite{Pratt:2003ar} in obtaining the
kernels for strong interactions, e.g., the kernels in
Fig.~\ref{fig:kernel_pkplus}. Energy derivatives additionally
constrain the integrals of the wavefunctions $R_\ell$ squared over
the region $r < d$ \cite{PhysRevC.33.1303,Pratt:2003ar}. This
facilitates an extrapolation of the kernels down to $r=0$,
adequate for source resolutions that can be practically achieved,
without the need to resort to potential models for the strong
interaction. Still, with the strong-interaction kernels becoming
model-dependent at short distances, we refrain from displaying
them at $r < 1.5 \, \text{fm}$.  Needless to say that, given the
quark substructure of hadrons, the description in terms of
wavefunction for relative motion becomes questionable at the
shortest distances $r \lesssim 0.5 \, \text{fm}$.  For IMFs the
same becomes true at even larger distances, due to the finite size of the nuclei.

\section{Spherical Cartesian Harmonics}
\label{sec:cartesian}

\subsection{Harmonics within Spherical Coordinates}

We start out with a review of basic concepts concerning surface spherical harmonics.
A harmonic function is one which satisfies the Laplace equation
\begin{equation}\label{eq:Laplace}
  \nabla^2 \, F({\pmb r}) = 0 \, .
\end{equation}
When using spherical coordinates,
it is convenient to express harmonic functions, which are regular at the origin,
in terms of tesseral harmonics $Y_{\ell m}$,
\begin{equation}\label{eq:Flm}
  F({\pmb r}) = \sqrt{4 \pi} \sum_{\ell m} F_{\ell m}^* \, r^\ell \, Y_{\ell m} (\Omega ) \, ,
\end{equation}
where the coefficients $F_{\ell m}$ take on arbitrary values.
In the summation over $\ell$, the individual terms in (\ref{eq:Flm}),
of the form
\begin{equation}\label{eq:Fl}
  F^{(\ell)} ({\pmb r}) = \sqrt{4 \pi} \, r^\ell  \sum_{m} F_{\ell m}^* \, Y_{\ell m} (\Omega ) \, ,
\end{equation}
are harmonic functions of degree~$\ell$.  The latter stand for
homogenous functions of degree $\ell$ in $x$, $y$ and $z$,
that are harmonic.
A surface spherical of degree $\ell$ is a harmonic function of degree $\ell$, taken on a unit sphere, $r=1$:
\begin{equation}\label{eq:Fl1}
  F^{(\ell)} (\Omega ) = \sqrt{4 \pi} \sum_{m} F_{\ell m}^* \, Y_{\ell m} (\Omega ) \, .
\end{equation}

Given the completeness relation for the tesseral spherical harmonics,
\begin{equation}\label{eq:dYlm}
  \delta ( \Omega - \Omega ') = \frac{1}{4 \pi} \sum_\ell (2\ell + 1) \, P_\ell ({\pmb n} \cdot {\pmb n}')
  =  \sum_{\ell m} Y_{\ell m}^* (\Omega' ) \, Y_{\ell m} (\Omega ) \, ,
\end{equation}
any function of spherical angle, and in particular ${\mathcal R}$ and ${\mathcal S}$
in Eqs.\ (\ref{eq:cartesianmaster_exp})-(\ref{eq:cartesianmaster}), can be expanded in the tesseral
harmonics upon using (\ref{eq:dYlm}) with the identity
\begin{equation}\label{eq:FeqF}
  G(\Omega ) = \int {\rm d} \Omega' \, \delta (\Omega - \Omega') \, G(\Omega') \, .
\end{equation}
Equation (\ref{eq:Flm}) for the
harmonic functions represents an example of the expansion,
where the expansion coefficients have a specific radial dependence.
Within every rank $\ell$ of an expansion,
the expanded functions are generally described in terms of $(2 \ell +1)$ complex
coefficients $F_{\ell m}$.  However, for real expanded functions, the coefficients $F_{\ell,m=0}$ must be real
and, otherwise, the coefficients must satisfy $F_{\ell \, -m} = (-1)^m \, F_{\ell m}^*$.  Thus,
for each~$\ell$,
the expansion is described in terms $(2 \ell +1)$ independent real numbers.
Under rotations of the coordinate system, for a given $\ell$, the tesseral functions transform like
the components of a~spherical tensor of rank $\ell$.
For a function
that is independent of the coordinate choice,
the expansion coefficients
transform further in this fashion;
the sums over $m$ in (\ref{eq:Flm}), (\ref{eq:Fl}) and (\ref{eq:Rylm}) represent scalar products
of the spherical tensors.  In fact, the r.h.s.\ of Eq.\ (\ref{eq:dYlm})
also represents a superposition of the scalar products
and, in this, represents a covariant generalization of
the middle result in (\ref{eq:dYlm}), which is seen as a superposition of the scalar products
of the tensors in the case when the coordinate $z$-axis is oriented either along ${\pmb n}$ or~${\pmb n}'$.

\subsection{Cartesian Harmonics}
Cartesian harmonics allow functions of the spherical angle to be expressed as
superpositions of the scalar products
of cartesian tensors.  For real functions, the coefficients of expansion in
cartesian harmonics are real,
which is not necessarily the case when expanding functions in tesseral
harmonics. The expansion coefficients for cartesian harmonics are generally
easier to interpret than those for
tesseral harmonics.

Let us consider a homogenous function of degree $\ell$.  When using
cartesian coordinates, a homogenous function may be represented as
\begin{equation}\label{eq:fhomol}
  F^{(\ell)} ({\pmb r}) =
  \sum_{\alpha_1 \, \alpha_2 \ldots \alpha_\ell} F_{\alpha_1 \, \alpha_2 \ldots \alpha_\ell}^{(\ell)}
  r_{\alpha_1} \, r_{\alpha_2} \ldots r_{\alpha_\ell} \, ,
\end{equation}
where $\alpha=x,y,z$ are cartesian indices.  The following remarks can be made regarding this representation
in terms of cartesian coordinates.
Since the product of cartesian coordinates on the r.h.s.\ of (\ref{eq:fhomol})
is symmetric under the interchange of its terms, attention may be limited to the coefficients of
expansion $F_{\alpha_1 \, \alpha_2 \ldots \alpha_\ell}$
which are symmetric under the interchange of
indices.  Since the product of $\ell$ coordinates transforms further as a cartesian tensor of rank $\ell$,
the coefficients
$F_{\alpha_1 \, \alpha_2 \ldots \alpha_\ell}$ must transform as a cartesian tensor of rank $\ell$
for a function
$F^{(\ell)}$ that is independent of the choice of directions for coordinate axes.  Given
the symmetry of the tensorial coefficients, the r.h.s.\ of (\ref{eq:fhomol})
can be conveniently rewritten by grouping terms with different
numbers, $\ell_x$, $\ell_y$ and $\ell_z$, respectively,
of the $x$, $y$ and $z$ coefficients, obtaining the representation such as in (\ref{eq:cartesianmaster_exp}),
\begin{equation}\label{eq:fl}
  F^{(\ell)} ({\pmb r}) = \sum_{\vec{\ell}} \frac{\ell!}{\ell_x! \, \ell_y! \, \ell_z!} \,
  F_{\vec{\ell}} \, x^{\ell_x} \, y^{\ell_y} \, z^{\ell_z} \equiv F^\top \, {\pmb r}^\ell \, ,
\end{equation}
where $\ell = \ell_x + \ell_y +\ell_z$.  The last expression in (\ref{eq:fl}) is a schematic representation
for the convolution of the cartesian tensors.

\begin{figure}
\centerline{\includegraphics[width=0.5\textwidth]{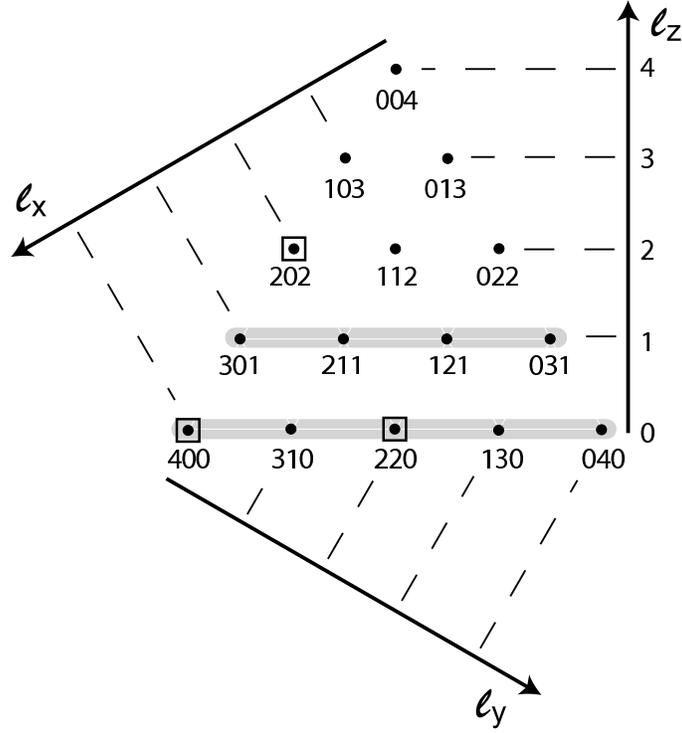}}
\caption{\label{fig:triangle}
Triangular diagram representing the components of a symmetric tensor, in terms of dots
labelled with the component $[\ell_x,\ell_y,\ell_z]$
values, for $\ell=\ell_x+\ell_y+\ell_z=4$, after \protect\cite{applequist}.
When moving along the lines parallel to the respective sides of the
triangle, either $\ell_x$, $\ell_y$ or $\ell_z$ stays constant.  The tracelessness condition
(\protect\ref{eq:Fnotracevec}) relates tensor components at the corners
of an equilateral triangle two $\ell$-units on the side, within the diagram, such
as those marked by the squares.  As the set of linearly independent components of a traceless symmetric tensor,
the $\ell_z = 0$ and $\ell_z=1$ components may be chosen, marked with the thick grey lines in the figure.
}
\end{figure}

On substituting the representation (\ref{eq:fhomol}) into the Laplace equation (\ref{eq:Laplace}), we
find that the homogenous function is harmonic
if and only if the cartesian tensorial coefficients are traceless \cite{applequist89},
i.e.,
\begin{equation}\label{eq:Fnotrace}
  \sum_{\alpha} F_{\alpha_1 \ldots \alpha_{\ell -2} \, \alpha \, \alpha} ^{(\ell)} = 0 \, ,
\end{equation}
or
\begin{equation}\label{eq:Fnotracevec}
  F_{\ell_x+2 , \ell_y , \ell_z} + F_{\ell_x , \ell_y+2 , \ell_z} + F_{\ell_x , \ell_y , \ell_z+2} = 0 \, ,
\end{equation}
where $\ell_x + \ell_y +\ell_z+2 = \ell $.
The components for a symmetric tensor, labelled with the values of $(\ell_x,\ell_y,\ell_z)$, are illustrated with a
triangle in Fig.~\ref{fig:triangle}.  When moving along lines parallel to the different sides of the triangle, either
the value of $\ell_x$, $\ell_y$ or $\ell_z$ stays constant.  The tracelessness condition (\ref{eq:Fnotracevec})
relates components at the corners of an equilateral triangle two $\ell$-units on the side in the figure.

In the appendix we show that there exists a real projection
operator $\mathcal P$, which operates in
the space of cartesian tensors of rank $\ell$ and projects, out
of any tensor, the part which is symmetric and traceless.  As an operator in the space of rank $\ell$ tensors,
$\mathcal P$ itself is a rank $2\ell$ tensor, $\ell$ times covariant and $\ell$ times contravariant.
As a projection operator, $ {\mathcal P}$
satisfies ${\mathcal P}^2 = {\mathcal P}$ and ${\mathcal P}^\top = {\mathcal P}$.
In terms of the operator $\mathcal P$, for any tensor $T$ of rank $\ell$, the
function given by
\begin{equation}\label{eq:TtoA}
  ({\mathcal P} T)^\top \, {\pmb r}^\ell = T^\top \, {\mathcal P}^\top \, {\pmb r}^\ell
  = T^\top \, {\mathcal P} \, {\pmb r}^\ell
\end{equation}
is harmonic, where we follow the schematic notation.
Since, the function (\ref{eq:TtoA}) is harmonic for any tensor $T$, each of the components of
${\mathcal P} \, {\pmb r}^\ell$ on the r.h.s.\ must be harmonic.
On the sphere of $r=1$, those components define the cartesian
surface spherical harmonics
\begin{equation}\label{eq:Adef}
  {\mathcal A}_{\alpha_1 \, \alpha_2 \ldots \alpha_\ell}^{(\ell)} (\Omega ) =
  \sum_{\alpha_1' \, \alpha_2 ' \ldots \alpha_\ell'}
  {\mathcal P}_{\alpha_1 \, \alpha_2  \ldots \alpha_\ell: \alpha_1 ' \, \alpha_2 ' \ldots \alpha_\ell '}^{(\ell : \ell)} \,
{n}_{\alpha_1 '} \, {n}_{\alpha_2 '} \ldots {n}_{\alpha_\ell '} \, ,
\end{equation}
where ${\pmb n}=(\sin{\theta} \cos{\phi}, \sin{\theta} \sin{\phi}, \cos{\theta})$,
which can be further also written as
\begin{equation}\label{eq:Adefvect}
  {\mathcal A}_{\vec{\ell}} (\Omega) = \sum_{\vec{\ell}'} \frac{\ell!}{\ell_x' ! \, \ell_y'! \, \ell_z'!} \,
  {\mathcal P}_{\vec{\ell}:\vec{\ell}'} \, n_x^{\ell_x'} \, n_y^{\ell_y'} \, n_z^{\ell_z'} \, .
\end{equation}

Cartesian components of the direction vector ${\pmb n}$
are combinations of the tesseral rank-1 harmonics $Y_{1m}$.
The product of $\ell$ components, as e.g.\ demonstrated by the quantal
rules of angular-momentum superposition, is a superposition
of tesseral harmonics of rank $\ell$ and lower, of the same evenness as $\ell$:
\begin{equation}\label{eq:AYlmcoarse}
  n_{\alpha_1} \, n_{\alpha_2} \ldots n_{\alpha_\ell} =
  \sum_{\substack{\ell' \, m'\\ \ell' \le \ell}}
  c_{\alpha_1 \, \alpha_2 \ldots \alpha_\ell}^{\ell' m' *} \, Y_{\ell' m'} (\Omega)
\equiv
{\mathcal A}_{\alpha_1 \, \alpha_2 \ldots \alpha_\ell}^{(\ell)} + \left(
n_{\alpha_1} \, n_{\alpha_2} \ldots n_{\alpha_\ell}
-  {\mathcal A}_{\alpha_1 \, \alpha_2 \ldots \alpha_\ell}^{(\ell)} \right)
  \, .
\end{equation}
Since $r^\ell {\mathcal A}^{(\ell)}$ are harmonic,
${\mathcal A}^{(\ell)}$ can consist of the tesseral rank-$\ell$ harmonics only.
As will be demonstrated in the appendix, the operator ${\mathcal P}$ consists out
of a symmetrization operator, i.e.\ the unit tensor in the space of symmetric cartesian tensors,
and of operator terms which involve taking traces of one or more pairs of the cartesian indices.
In acting on the the cartesian product ${\pmb n}^\ell$, the symmetrization operator leaves
the product ${\pmb n}^\ell$ intact while the other operator terms in ${\mathcal P}$
produce symmetrized terms of the form
${\pmb n}^{\ell - 2k}\, (\delta)^k$, where $k=1, \ldots, \ell/2$.  The latter terms represent
superpositions of tesseral harmonics up to the rank of $\ell-2$,
of the same evenness as $\ell$.  The implication is that the projection operator ${\mathcal P}$ removes
from ${\pmb n}^\ell$ a portion represented by the superposition
of the lower rank tesseral harmonics (in parenthesis
on the r.h.s.\ of (\ref{eq:AYlmcoarse})) while retaining the highest rank $\ell$ portion intact.
Given the general form of ${\mathcal A}^{(\ell)}$, from application of ${\mathcal P}$ onto ${\pmb n}^\ell$,
the cartesian harmonics may be constructed recursively \cite{Danielewicz:2005qh},
based on the tracelessness of the ${\mathcal A}^{(\ell)}$ tensors within each
rank, starting with ${\mathcal A}^{(\ell=0)}=1$:
\begin{eqnarray}
{\mathcal A}_{\alpha_1 \cdots \alpha_{\ell}}^{(\ell)} & = & \frac{1}{\ell}
\sum_{i=1}^{\ell} n_{\alpha_i} \, {\mathcal A}_{\alpha_1 \cdots \alpha_{i-1} \, \alpha_{i+1}
\cdots \alpha_{\ell}}^{(\ell - 1)}
\nonumber
\\
&& - \frac{2}{\ell \, (2\ell-1)}\sum_{1\le i < j \le \ell}
\sum_{\alpha} \delta_{\alpha_i \,\alpha_j} \,
n_\alpha \,
{\mathcal A}^{(\ell-1)}_{\alpha \,\alpha_1 \cdots \alpha_{i-1} \, \alpha_{i+1} \cdots
\alpha_{j-1} \, \alpha_{j+1} \cdots \alpha_{\ell}} \, .
\label{eq:Arecursion}
\end{eqnarray}
Expressions for cartesian harmonics of lowest rank are given in Table \ref{table:cartesian}
and the recursion with (\ref{eq:Arecursion}), otherwise, produces the series (\ref{eq:Anseries}).
One of the features of the cartesian harmonics is that the three cartesian axes are treated equally,
unlike in the case of tesseral harmonics.  Within a given rank $\ell$, not all cartesian harmonics are independent.
A symmetric tensor has $(\ell+1) (\ell +2)/2$ different components, cf.\ Fig.\ \ref{fig:triangle}.
The tracelessness condition (\ref{eq:Fnotracevec}) reduces the number of linearly independent components, including
the number of independent ${\mathcal A}^{(\ell)}$ functions, to $(2 \ell + 1)$, equal to the number of
$Y_{\ell m}$ functions.  As a linearly independent set, one might choose the components with $\ell_z=0$ and
$\ell_z=1$, see the figure.

\begin{table}
\caption{\label{table:cartesian} Cartesian harmonics for $\ell \le 4$. Other
harmonics can be found by either permuting the indices, i.e.\ ${\mathcal
A}_{xyx}={\mathcal A}_{xxy}$, or by swapping indices on both sides of an
equality, e.g.\ $x\leftrightarrow y$.  Thus, given ${\mathcal
A}_{xxy}=n_x^2 \, n_y - (1/5) n_y$, swapping the indices $y\leftrightarrow z$ yields ${\mathcal
A}_{xxz}=n_x^2 \, n_z- (1/5) n_z$.}
\begin{tabular}{|c|c|}\hline
${\mathcal A}^{(1)}_x=n_x$
& ${\mathcal A}^{(3)}_{xyz}=n_x \, n_y \, n_z$ \\
${\mathcal A}^{(2)}_{xx}=n_x^2- 1/3$
& $A^{(4)}_{xxxx}=n_x^4-(6/7)n_x^2+3/35$ \\
${\mathcal A}^{(2)}_{xy}=n_x \, n_y$
& ${\mathcal A}^{(4)}_{xxxy}=n_x^3 \, n_y-(3/7)n_x \, n_y$\\
$A^{(3)}_{xxx}=n_x^3-(3/5)n_x$
&  ${\mathcal A}^{(4)}_{xxyy}=n_x^2 \, n_y^2-(1/7)n_x^2-(1/7)n_y^2+1/35$\\
${\mathcal A}^{(3)}_{xxy}=n_x^2 \, n_y-(1/5)n_y$
& ${\mathcal A}^{(4)}_{xxyz}=n_x^2 \, n_y \, n_z-(1/7)n_y \, n_z$\\ \hline
\end{tabular}
\end{table}

In the case of $\alpha_1 = \alpha_2 = \ldots = \alpha_\ell = z$ in (\ref{eq:AYlmcoarse}), both
the cartesian component product ${\pmb n}^\ell$
and ${\mathcal A}^{(\ell)}$ are invariant with respect to rotations
around the $z$ axis.  For invariance, both functions must be a superposition of $m'=0$ tesseral harmonics only.
It follows then that ${\mathcal A}_{z \ldots z}^{(\ell)}$ must be proportional to $Y_{\ell 0}$ or $P_\ell$.
Since the action of the projection operator ${\mathcal P}$ leaves the highest rank terms unchanged, the
coefficient for $P_\ell$ in ${\mathcal A}^{(\ell)}$ must be the same as in ${\pmb n}^\ell$.  Accordingly,
the respective cartesian harmonics are given by
\begin{equation}\label{eq:Azz}
  {\mathcal A}_{z \, z \ldots z}^{(\ell)} (\Omega)
  = \frac{\ell !} {(2 \ell -1)!!} \, P_\ell (\cos{\theta}) \, .
\end{equation}
Starting with the first equality in (\ref{eq:dYlm}), we are now in position to
produce a completeness relation in terms
of the cartesian harmonics.  Thus, if ${\pmb n}'$ is directed along the $z$-axis in (\ref{eq:dYlm}),
we can express, given (\ref{eq:Azz}), the Legendre polynomial in the middle term in (\ref{eq:dYlm}) as
$((2 \ell -1)!!/\ell !) ({\pmb n}^{\prime \ell})^\top \, {\mathcal A}$.  When rotating ${\pmb n}'$
and ${\pmb n}$, to move ${\pmb n}'$ away from the alignment with the $z$-axis, the scalar product
of the tensors transforms covariantly, allowing to obtain different general representations for
the $\delta$-function
in spherical angle in terms of the cartesian harmonics
\begin{eqnarray}
  \delta ( \Omega - \Omega ') & = & \frac{1}{4 \pi} \sum_\ell (2\ell + 1) \, P_\ell ({\pmb n} \cdot {\pmb n}')
  \nonumber \\
  & = & \frac{1}{4 \pi} \sum_\ell \frac{(2 \ell + 1)!!} {\ell !} \sum_{\alpha_1 \ldots \alpha_\ell}
  n_{\alpha_1}' \ldots n_{\alpha_\ell}' \, {\mathcal A}_{\alpha_1 \ldots \alpha_\ell}^{(\ell)} (\Omega)
  \nonumber \\
  & = & \frac{1}{4 \pi} \sum_\ell \frac{(2 \ell + 1)!!} {\ell !} \sum_{\alpha_1 \ldots \alpha_\ell}
  {\mathcal A}_{\alpha_1 \ldots \alpha_\ell}^{(\ell)} (\Omega ') \,
  {\mathcal A}_{\alpha_1 \ldots \alpha_\ell}^{(\ell)} (\Omega)
   \nonumber \\
  & = & \frac{1}{4 \pi} \sum_{\vec{\ell}} \frac{(2 \ell + 1)!!} {\ell_x! \, \ell_y! \, \ell_z!} \,
  {\mathcal A}_{\vec{\ell}} (\Omega ') \, {\mathcal A}_{\vec{\ell}} (\Omega )
  \nonumber \\
  & = & \frac{1}{4 \pi} \sum_{\vec{\ell}} \frac{(2 \ell + 1)!!} {\ell_x! \, \ell_y! \, \ell_z!} \,
  {\mathcal A}_{\vec{\ell}} (\Omega ') \, n_x^{\ell_x} \, n_y^{\ell_y} \, n_z^{\ell_z} \, .
\label{eq:dA}
\end{eqnarray}
In obtaining the third equality in (\ref{eq:dA}) and later, we use the projection properties of the
operator ${\mathcal P}$:
\begin{equation}\label{eq:nPA}
  ({\pmb n}^{\prime \ell})^\top \, {\mathcal A} = ({\pmb n}^{\prime \ell})^\top \, {\mathcal P} \, {\mathcal A}
  = ( {\mathcal P} \, {\pmb n}^{\prime \ell})^\top \, {\mathcal A} = ({\mathcal A}')^\top \, {\mathcal A} \, .
\end{equation}
Use of (\ref{eq:dA}) in the identity (\ref{eq:FeqF})
allows for functions of spherical angle to be expressed as a sum of
cartesian harmonics, such as in (\ref{eq:cartesianmaster_exp}).  Independent of the orientation of the coordinate axes, the coefficients of expansion in cartesian harmonics,
such as in (\ref{eq:cartesianmaster_coef}),
transform as components of a cartesian tensor
under rotations.  As the tensor is traceless, only $(2 \ell+1)$ coefficients are independent within each rank~$\ell$.

\subsection{Operations with Cartesian Harmonics}

When given some specific expansion coefficients in terms of cartesian harmonics,
it may be of interest to calculate the associated cartesian moments.
To facilitate that, it is useful to express products of cartesian components in terms of cartesian
harmonics, which can be done \cite{Danielewicz:2005qh} through recursion
starting from Eq.~(\ref{eq:Anseries}), obtaining
\begin{eqnarray}
\nonumber
n_x^{\ell_x} \, n_y^{\ell_y} \, n_z^{\ell_z}
& = &
\sum_{\substack{ \vec{m} \\ 0 \le m_i \le \ell_i/2}}
\frac{(2\ell-4m+1)!!}{2^m \, (2\ell-2m+1)!!} \\
&& \hspace*{2em} \times \frac{\ell_x!}{(\ell_x-2m_x)! \, m_x!} \,
\frac{\ell_y!}{(\ell_y-2m_y)! \, m_y!} \,
\frac{\ell_z!}{(\ell_z-2m_z)! \, m_z!} \,
{\mathcal A}_{\vec{\ell}-2\vec{m}}(\Omega)  \, .
\label{eq:nexpansion}
\end{eqnarray}
The rank of ${\mathcal A}$ within each cartesian component on the r.h.s.\ is equal to or lower than the
power of the corresponding cartesian component on the l.h.s.  From (\ref{eq:nexpansion})
and (\ref{eq:cartesianmaster_coef}), we find
\begin{eqnarray}
\label{eq:cartesiansourcemoments}
\int {\rm d}^3 r \, {\mathcal S}({\pmb r}) \, x^{\ell_x} \, y^{\ell_y} \, z^{\ell_z}
& = &
\frac{4 \pi \, \ell!}{(2\ell+1)!!}
\sum_{\substack{ \vec{m} \\ 0 \le m_i \le \ell_i/2}}
\frac{(2\ell-4m+1)!!}{2^m \, (2\ell-2m+1)!!}\\
\nonumber
&& \hspace*{2em} \times \frac{\ell_x!}{(\ell_x-2m_x)! \, m_x!} \,
\frac{\ell_y!}{(\ell_y-2m_y)! \, m_y!} \,
\frac{\ell_z!}{(\ell_z-2m_z)! \, m_z!} \,
\int_0^\infty {\rm d}r \, r^{\ell+2} \,
{\mathcal S}_{\vec{\ell}-2\vec{m}}(r) \, ,
\end{eqnarray}
which allows the moments to be obtained from cartesian harmonic coefficients.

As combinations of tesseral harmonics of different rank~$\ell$, the cartesian harmonics
of different rank are orthogonal to one another.  Due to different behavior under inversion, see
Eq.~(\ref{eq:Anseries}), two cartesian harmonics are further orthogonal if any pair of
their $\ell_i$ subscripts
differs by an odd number.  Otherwise, the scalar product of harmonics is given by \cite{applequist}
\begin{eqnarray}
\nonumber
\langle \vec{\ell}|\vec{\ell}' \rangle \equiv \int \frac{{\rm d} \Omega}{4 \pi} \, {\mathcal A}_{\vec{\ell}} \,
{\mathcal A}_{\vec{\ell}'} & = &
\frac{1}{(2\ell+1)!! \, (2\ell-1)!!}
\sum_{ \substack{ \vec{m} \\ \max{(0,(\ell_i' - \ell_i)/2)} \le m_i \\ m_i
\le \min{(\ell_i' - \ell_i/2, \ell_i/2 )}}}
\left(- \frac{1}{2}\right)^m \, m! \, (2\ell-2m-1)!!\\
&&\times \prod_{i=x,y,z} \frac{\ell_i! \, \ell'_i!}{((\ell'_i-\ell_i)/2+m_i)! \, (\ell_i-2m_i)! \, m_i!}
\, .
\label{eq:overlap}
\end{eqnarray}
The anticipated symmetry between $\vec{\ell}$ and $\vec{\ell}'$ on the r.h.s. of (\ref{eq:overlap})
may be seen by eliminating
$m_i$ in favor of $m_i' = m_i + (\ell_i - \ell_i')/2$.

Cartesian expansion coefficients may be converted into tesseral expansion coefficients and {\em vice versa}.
In terms of cartesian harmonics, the $m \ge 0$ tesseral harmonics are given by
 \cite{applequist}
\begin{eqnarray}
Y_{\ell m} (\Omega ) & = & (-1)^{m} \, (2\ell-1)!!
\left[\frac{2\ell+1}{4\pi \, (\ell+m)! \, (\ell-m)!}\right]^{1/2} \,
\sum_{k=0}^m i^{m-k} \, \frac{m!}{(m-k)! \, k!} \,
{\mathcal A}_{[k,m-k,\ell-m]}(\Omega)  \, .
\end{eqnarray}
The cartesian harmonics may be, on the other hand, expressed as
\begin{equation}\label{eq:A=Y}
{\mathcal A}_{\vec{\ell}} (\Omega)
= \sum_{m=-\ell}^{\ell} \big(a_{\vec{\ell}}^{m}\big)^* \, Y_{\ell m} (\Omega) \, ,
\end{equation}
where \cite{applequist}, for $m \ge 0$,
\begin{equation}\label{eq:C=}
  a_{\vec{\ell}}^{m} = (-1)^m \, (2\ell-1)!!
  \left[\frac{2\ell+1}{4\pi \, (\ell+m)! \, (\ell-m)!}\right]^{1/2} \,
  \sum_{k=0}^m i^{m-k} \, \frac{m!}{(m-k)! \,  k!} \,
  \big\langle \vec{\ell} \big| [k,m-k,\ell - m] \big\rangle \, ,
\end{equation}
the overlap integral $\langle \cdot | \cdot \rangle$ is given in (\ref{eq:overlap}) and
$a_{\vec{\ell}}^{-m}= (-1)^m \, \big(a_{\vec{\ell}}^{m}\big)^*$.
Formulas for conversion between tesseral and cartesian harmonics for $\ell \le 4$
are given in Table \ref{table:YofA}.

 \begin{table}
\caption{\label{table:YofA}
Transformation equations between cartesian and tesseral harmonics for $\ell\le 4$.}
\begin{tabular}{|l|}\hline
$Y_{1 \, 0} =
(1/2)\sqrt{3/\pi} \, {\mathcal A}^{(1)}_{z}$\\
$Y_{1 \, \pm 1} =
 -(i/2)\sqrt{3/2\pi} \, {\mathcal A}^{(1)}_{y}
 \mp(1/2)\sqrt{3/2\pi} \, {\mathcal A}^{(1)}_{x}$\\
$Y_{2 \, 0} =
(3/4)\sqrt{5/\pi} \, {\mathcal A}^{(2)}_{zz}$\\
$Y_{2 \, \pm 1} =
 -(i/2)\sqrt{15/2\pi} \, {\mathcal A}^{(2)}_{yz}
 \mp(1/2)\sqrt{15/2\pi} \, {\mathcal A}^{(2)}_{xz}$\\
$Y_{2 \, \pm 2} =
 -(1/4)\sqrt{15/2\pi} \, {\mathcal A}^{(2)}_{yy}
 \pm(i/2)\sqrt{15/2\pi} \, {\mathcal A}^{(2)}_{xy}
 +(1/4)\sqrt{15/2\pi} \, {\mathcal A}^{(2)}_{xx}$\\
$Y_{3 \, 0} =
(5/4)\sqrt{7/\pi} \, {\mathcal A}^{(3)}_{zzz}$\\
$Y_{3 \, \pm 1} =
 -(5i/8)\sqrt{21/\pi} \, {\mathcal A}^{(3)}_{yzz}
 \mp(5/8)\sqrt{21/\pi} \, {\mathcal A}^{(3)}_{xzz}$\\
$Y_{3 \, \pm 2} =
 -(1/4)\sqrt{105/2\pi} \, {\mathcal A}^{(3)}_{yyz}
 \pm(i/2)\sqrt{105/2\pi} \, {\mathcal A}^{(3)}_{xyz}
 +(1/4)\sqrt{105/2\pi} \, {\mathcal A}^{(3)}_{xxz}$\\
$Y_{3 \, \pm 3} =
(i/8)\sqrt{35/\pi} \, {\mathcal A}^{(3)}_{yyy}
 \pm(3/8)\sqrt{35/\pi} \, {\mathcal A}^{(3)}_{xyy}
 -(3i/8)\sqrt{35/\pi} \, {\mathcal A}^{(3)}_{xxy}
 \mp(1/8)\sqrt{35/\pi} \, {\mathcal A}^{(3)}_{xxx}$\\
$Y_{4 \, 0} =
(105/16)\sqrt{1/\pi} \, {\mathcal A}^{(4)}_{zzzz}$\\
$Y_{4 \, \pm 1} =
 -(21i/8)\sqrt{5/\pi} \, {\mathcal A}^{(4)}_{yzzz}
 \mp(21/8)\sqrt{5/\pi} \, {\mathcal A}^{(4)}_{xzzz}$\\
$Y_{4 \, \pm 2} =
 -(21/8)\sqrt{5/2\pi} \, {\mathcal A}^{(4)}_{yyzz}
 \pm(21i/4)\sqrt{5/2\pi} \, {\mathcal A}^{(4)}_{xyzz}
 +(21/8)\sqrt{5/2\pi} \, {\mathcal A}^{(4)}_{xxzz}$\\
$Y_{4 \, \pm 3} =
(3i/8)\sqrt{35/\pi} \, {\mathcal A}^{(4)}_{yyyz}
 \pm(9/8)\sqrt{35/\pi} \, {\mathcal A}^{(4)}_{xyyz}
 -(9i/8)\sqrt{35/\pi} \, {\mathcal A}^{(4)}_{xxyz}
 \mp(3/8)\sqrt{35/\pi} \, {\mathcal A}^{(4)}_{xxxz}$\\
$Y_{4 \, \pm 4} =
(3/16)\sqrt{35/2\pi} \, {\mathcal A}^{(4)}_{yyyy}
 \mp(3i/4)\sqrt{35/2\pi} \, {\mathcal A}^{(4)}_{xyyy}
 -(9/8)\sqrt{35/2\pi} \, {\mathcal A}^{(4)}_{xxyy}
 \pm(3i/4)\sqrt{35/2\pi} \, {\mathcal A}^{(4)}_{xxxy}
 +(3/16)\sqrt{35/2\pi} \, {\mathcal A}^{(4)}_{xxxx}$\\
\hline
${\mathcal A}^{(1)}_{z} =
2\sqrt{\pi/3} \, Y_{1 \, 0}$\\
${\mathcal A}^{(1)}_{y} =
i\sqrt{2\pi/3} \, Y_{1 \, 1}
 +i\sqrt{2\pi/3} \, Y_{1 \, -1}$\\
${\mathcal A}^{(1)}_{x} =
 -\sqrt{2\pi/3} \, Y_{1 \, 1}
 +\sqrt{2\pi/3} \, Y_{1 \, -1}$\\
${\mathcal A}^{(2)}_{zz} =
(4/3)\sqrt{\pi/5} \, Y_{2 \, 0}$\\
${\mathcal A}^{(2)}_{yz} =
i\sqrt{2\pi/15} \, Y_{2 \, 1}
 +i\sqrt{2\pi/15} \, Y_{2 \, -1}$\\
${\mathcal A}^{(2)}_{yy} =
 -\sqrt{2\pi/15} \, Y_{2 \, 2}
 -(2/3)\sqrt{\pi/5} \, Y_{2 \, 0}
 -\sqrt{2\pi/15} \, Y_{2 \, -2}$\\
${\mathcal A}^{(2)}_{xz} =
 -\sqrt{2\pi/15} \, Y_{2 \, 1}
 +\sqrt{2\pi/15} \, Y_{2 \, -1}$\\
${\mathcal A}^{(2)}_{xy} =
 -i\sqrt{2\pi/15} \, Y_{2 \, 2}
 +i\sqrt{2\pi/15} \, Y_{2 \, -2}$\\
${\mathcal A}^{(2)}_{xx} =
\sqrt{2\pi/15} \, Y_{2 \, 2}
 -(2/3)\sqrt{\pi/5} \, Y_{2 \, 0}
 +\sqrt{2\pi/15} \, Y_{2 \, -2}$\\
${\mathcal A}^{(3)}_{zzz} =
(4/5)\sqrt{\pi/7} \, Y_{3 \, 0}$\\
${\mathcal A}^{(3)}_{yzz} =
(4i/5)\sqrt{\pi/21} \, Y_{3 \, 1}
 +(4i/5)\sqrt{\pi/21} \, Y_{3 \, -1}$\\
${\mathcal A}^{(3)}_{yyz} =
 -\sqrt{2\pi/105} \, Y_{3 \, 2}
 -(2/5)\sqrt{\pi/7} \, Y_{3 \, 0}
 -\sqrt{2\pi/105} \, Y_{3 \, -2}$\\
${\mathcal A}^{(3)}_{yyy} =
 -i\sqrt{\pi/35} \, Y_{3 \, 3}
 -(i/5)\sqrt{3\pi/7} \, Y_{3 \, 1}
 -(i/5)\sqrt{3\pi/7} \, Y_{3 \, -1}
 -i\sqrt{\pi/35} \, Y_{3 \, -3}$\\
${\mathcal A}^{(3)}_{xzz} =
 -(4/5)\sqrt{\pi/21} \, Y_{3 \, 1}
 +(4/5)\sqrt{\pi/21} \, Y_{3 \, -1}$\\
${\mathcal A}^{(3)}_{xyz} =
 -i\sqrt{2\pi/105} \, Y_{3 \, 2}
 +i\sqrt{2\pi/105} \, Y_{3 \, -2}$\\
${\mathcal A}^{(3)}_{xyy} =
\sqrt{\pi/35} \, Y_{3 \, 3}
 +(1/5)\sqrt{\pi/21} \, Y_{3 \, 1}
 -(1/5)\sqrt{\pi/21} \, Y_{3 \, -1}
 -\sqrt{\pi/35} \, Y_{3 \, -3}$\\
${\mathcal A}^{(3)}_{xxz} =
\sqrt{2\pi/105} \, Y_{3 \, 2}
 -(2/5)\sqrt{\pi/7} \, Y_{3 \, 0}
 +\sqrt{2\pi/105} \, Y_{3 \, -2}$\\
${\mathcal A}^{(3)}_{xxy} =
i\sqrt{\pi/35} \, Y_{3 \, 3}
 -(i/5)\sqrt{\pi/21} \, Y_{3 \, 1}
 -(i/5)\sqrt{\pi/21} \, Y_{3 \, -1}
 +i\sqrt{\pi/35} \, Y_{3 \, -3}$\\
${\mathcal A}^{(3)}_{xxx} =
 -\sqrt{\pi/35} \, Y_{3 \, 3}
 +(1/5)\sqrt{3\pi/7} \, Y_{3 \, 1}
 -(1/5)\sqrt{3\pi/7} \, Y_{3 \, -1}
 +\sqrt{\pi/35} \, Y_{3 \, -3}$\\
${\mathcal A}^{(4)}_{zzzz} =
(16/105)\sqrt{\pi} \, Y_{4 \, 0}$\\
${\mathcal A}^{(4)}_{yzzz} =
(4i/21)\sqrt{\pi/5} \, Y_{4 \, 1}
 +(4i/21)\sqrt{\pi/5} \, Y_{4 \, -1}$\\
${\mathcal A}^{(4)}_{yyzz} =
 -(2/21)\sqrt{2\pi/5} \, Y_{4 \, 2}
 -(8/105)\sqrt{\pi} \, Y_{4 \, 0}
 -(2/21)\sqrt{2\pi/5} \, Y_{4 \, -2}$\\
${\mathcal A}^{(4)}_{yyyz} =
 -(i/3)\sqrt{\pi/35} \, Y_{4 \, 3}
 -(i/7)\sqrt{\pi/5} \, Y_{4 \, 1}
 -(i/7)\sqrt{\pi/5} \, Y_{4 \, -1}
 -(i/3)\sqrt{\pi/35} \, Y_{4 \, -3}$\\
${\mathcal A}^{(4)}_{yyyy} =
(1/3)\sqrt{2\pi/35} \, Y_{4 \, 4}
 +(2/21)\sqrt{2\pi/5} \, Y_{4 \, 2}
 +(2/35)\sqrt{\pi} \, Y_{4 \, 0}
 +(2/21)\sqrt{2\pi/5} \, Y_{4 \, -2}
 +(1/3)\sqrt{2\pi/35} \, Y_{4 \, -4}$\\
${\mathcal A}^{(4)}_{xzzz} =
 -(4/21)\sqrt{\pi/5} \, Y_{4 \, 1}
 +(4/21)\sqrt{\pi/5} \, Y_{4 \, -1}$\\
${\mathcal A}^{(4)}_{xyzz} =
 -(2i/21)\sqrt{2\pi/5} \, Y_{4 \, 2}
 +(2i/21)\sqrt{2\pi/5} \, Y_{4 \, -2}$\\
${\mathcal A}^{(4)}_{xyyz} =
(1/3)\sqrt{\pi/35} \, Y_{4 \, 3}
 +(1/21)\sqrt{\pi/5} \, Y_{4 \, 1}
 -(1/21)\sqrt{\pi/5} \, Y_{4 \, -1}
 -(1/3)\sqrt{\pi/35} \, Y_{4 \, -3}$\\
${\mathcal A}^{(4)}_{xyyy} =
(i/3)\sqrt{2\pi/35} \, Y_{4 \, 4}
 +(i/21)\sqrt{2\pi/5} \, Y_{4 \, 2}
 -(i/21)\sqrt{2\pi/5} \, Y_{4 \, -2}
 -(i/3)\sqrt{2\pi/35} \, Y_{4 \, -4}$\\
${\mathcal A}^{(4)}_{xxzz} =
(2/21)\sqrt{2\pi/5} \, Y_{4 \, 2}
 -(8/105)\sqrt{\pi} \, Y_{4 \, 0}
 +(2/21)\sqrt{2\pi/5} \, Y_{4 \, -2}$\\
${\mathcal A}^{(4)}_{xxyz} =
(i/3)\sqrt{\pi/35} \, Y_{4 \, 3}
 -(i/21)\sqrt{\pi/5} \, Y_{4 \, 1}
 -(i/21)\sqrt{\pi/5} \, Y_{4 \, -1}
 +(i/3)\sqrt{\pi/35} \, Y_{4 \, -3}$\\
${\mathcal A}^{(4)}_{xxyy} =
 -(1/3)\sqrt{2\pi/35} \, Y_{4 \, 4}
 +(2/105)\sqrt{\pi} \, Y_{4 \, 0}
 -(1/3)\sqrt{2\pi/35} \, Y_{4 \, -4}$\\
${\mathcal A}^{(4)}_{xxxz} =
 -(1/3)\sqrt{\pi/35} \, Y_{4 \, 3}
 +(1/7)\sqrt{\pi/5} \, Y_{4 \, 1}
 -(1/7)\sqrt{\pi/5} \, Y_{4 \, -1}
 +(1/3)\sqrt{\pi/35} \, Y_{4 \, -3}$\\
${\mathcal A}^{(4)}_{xxxy} =
 -(i/3)\sqrt{2\pi/35} \, Y_{4 \, 4}
 +(i/21)\sqrt{2\pi/5} \, Y_{4 \, 2}
 -(i/21)\sqrt{2\pi/5} \, Y_{4 \, -2}
 +(i/3)\sqrt{2\pi/35} \, Y_{4 \, -4}$\\
${\mathcal A}^{(4)}_{xxxx} =
(1/3)\sqrt{2\pi/35} \, Y_{4 \, 4}
 -(2/21)\sqrt{2\pi/5} \, Y_{4 \, 2}
 +(2/35)\sqrt{\pi} \, Y_{4 \, 0}
 -(2/21)\sqrt{2\pi/5} \, Y_{4 \, -2}
 +(1/3)\sqrt{2\pi/35} \, Y_{4 \, -4}$\\
\hline
\end{tabular}
\end{table}

Many circumstances require multiplication or division of the functions
of spherical angle.  For example, the empirical
correlation function $C=1+{\mathcal R}$, as a function of spherical angle at a given ${\pmb P}$ and $q$, follows
from an equation of the form
\begin{equation}
\label{eq:XBC}
X(\Omega)= C(\Omega) \, B(\Omega) \, ,
\end{equation}
where the l.h.s.\ represents the measured two-particle yield, $X(\Omega) \equiv \frac{{\rm d}^6N^{ab}}{{\rm d}^3p_a \,
{\rm d}^3p_b}$, cf.\ Eq.~(\ref{eq:corrmaster}), while $B$ on the r.h.s.\ is the product of single-particle yields,
$\frac{{\rm d} N^a}{{\rm d}^3p_a} \,
\frac{{\rm d}N^b}{{\rm d}^3p_b}$, and possibly of two-particle efficiency.  In order to find
$C (\Omega )$, one could bin the the three functions in two dimensions of $\cos\theta$ and $\phi$ and
obtain $C$ within each bin by dividing the value of $X$ in the bin by the value of $B$ in the bin.  However,
if one next wanted to expand $C$ in cartesian harmonics, the result could be distorted by
residual binning effects which would disappear in the limit of fine binning, but that could be only afforded
for high statistics.

An alternative approach is to explicitly solve for the coefficients
$C_{\vec{\ell}}$ given $X_{\vec{\ell}}$ and $B_{\vec{\ell}}$.
With this, one could forego the binning in $\cos{\theta}$ and $\phi$.
For instance, when assessing $X$ and identifying a particle pair at an angle $\Omega$,
one could increment values in an array for $X_{\vec{\ell}}$ by
$[(2\ell+1)!!/\ell!] \, {\mathcal A}_{\vec{\ell}}(\Omega)$ as described by Eq.~(\ref{eq:cartesianmaster_coef}).
One could then carry out an analogous procedure for the mixed pairs, filtering them through
two-particle efficiency.  In this way, both $X_{\vec{\ell}}$ and $B_{\vec{\ell}}$ could be obtained
in a straightforward fashion, without any binning in $\cos{\theta}$ and $\phi$.

Though the determination of $X_{\vec{\ell}}$ and $B_{\vec{\ell}}$ could be straightforward,
the determination of $C_{\vec{\ell}}$ can be more difficult.  Here, we outline one possible
procedure based on the expansion (\ref{eq:cartesianmaster_exp}) of the functions in powers of~$\hat{q}$.
If we introduce
\begin{equation}\label{eq:gamma=}
 \gamma(\vec{\ell})= \frac{\ell!}{\ell_x! \, \ell_y! \, \ell_z!} \, ,
\end{equation}
we can write
\begin{equation}
\label{eq:exexpansiona}
X(\hat{q}) =
\sum_{\vec{\ell}} \gamma(\vec{\ell}) \, X_{\vec{\ell}} \,
\hat{q}_x^{\ell_x}
\hat{q}_y^{\ell_y}
\hat{q}_z^{\ell_z}
= \sum_{\vec{\ell}} \sum_{\substack{ \vec{\ell}' \\ 0 \le \ell_i' \le \ell_i}}
\gamma(\vec{\ell}-\vec{\ell}') \,
\gamma(\vec{\ell}') \,
B_{\vec{\ell} - \vec{\ell}'} \, C_{\vec{\ell}'} \,
\hat{q}_x^{\ell_x} \,
\hat{q}_y^{\ell_y} \,
\hat{q}_z^{\ell_z} \, .
\end{equation}
If the condition of tracelessness for $\bar{C}_{\vec{\ell}}$
is lifted, the solution to Eq.~(\ref{eq:exexpansiona}) ceases to be unique, because
$\hat{q}_i$ are not independent
due to the constraint $\hat{q}_x^2+\hat{q}_y^2+\hat{q}_z^2=1$. We will first find a solution to (\ref{eq:exexpansiona})
without imposing the tracelessness condition,
which will provide us with $C(\Omega)$ from which we can next find the traceless~$C_{\vec{\ell}}$.

The simplest strategy in finding a solution to (\ref{eq:exexpansiona})
is to treat the components $\hat{q}_i$ as if they were
independent.  If we denote the solution by $\bar{C}_{\vec{\ell}}$, we obtain
\begin{equation}\label{eq:exexpansion}
  \gamma(\vec{\ell}) \, X_{\vec{\ell}} =  \sum_{\substack{ \vec{\ell}' \\ 0 \le \ell_i' \le \ell_i}}
  \gamma(\vec{\ell}-\vec{\ell}') \,
\gamma(\vec{\ell}') \,
B_{\vec{\ell} - \vec{\ell}'} \, \bar{C}_{\vec{\ell}'} \, ,
\end{equation}
which may be solved for $\bar{C}_{\vec{\ell}}$ iteratively.  The $\ell=0$ term follows right away,
\begin{equation}
\label{eq:C0=}
\bar{C}_{0}=\frac{X_{0}}{B_{0}} \, ,
\end{equation}
while the subsequent coefficients follow from
\begin{equation}\label{eq:Cl=}
  \bar{C}_{\vec{\ell}} = \frac{X_{\vec{\ell}}}{B_{\vec{\ell}}} -
  \frac{1}{\gamma(\vec{\ell}) \, B_{\vec{\ell}}}
  \sum_{\substack{\vec{\ell}' \\ \ell' \le \ell-1 \\ 0 \le \ell_i' \le \ell_i
   }} \gamma(\vec{\ell}-\vec{\ell}') \, \gamma(\vec{\ell}') \, B_{\vec{\ell} - \vec{\ell}'} \, \bar{C}_{\vec{\ell}'} \, .
\end{equation}
The combination of
\begin{equation}
\label{eq:COmega}
C (\Omega) = \sum_{\vec{\ell}} \gamma(\vec{\ell}) \, \bar{C}_{\vec{\ell}} \,
\hat{q}_x^{\ell_x} \, \hat{q}_y^{\ell_y} \, \hat{q}_z^{\ell_z} \, ,
\end{equation}
and of (\ref{eq:cartesianmaster_coef}) next yields for the traceless coefficients
\begin{equation}\label{eq:Cell}
  C_{\vec{\ell}} =
\frac{(2\ell+1)!!}{\ell!}\int \frac{d\Omega}{4\pi} \,
{\mathcal A}_{\vec{\ell}}(\Omega) \, C(\Omega )
= \frac{(2\ell+1)!!}{\ell!} \sum_{\vec{m}} \gamma (\vec{\ell}+ 2 \vec{m}) \, \langle \vec{\ell} |
\vec{\ell}+ 2 \vec{m}] \, \bar{C}_{\vec{\ell}+ 2 \vec{m}}
\, ,
\end{equation}
with
\begin{eqnarray}
  \langle \vec{\ell} | \vec{\ell}' ]  & \equiv &  \int \frac{{\rm d} \Omega}{4 \pi} \,
  {\mathcal A}_{\vec{\ell}} (\Omega) \,
  \hat{n}_x^{\ell_x'} \,  \hat{n}_y^{\ell_y'} \,  \hat{n}_z^{\ell_z'} \nonumber \\
  & = & \frac{ (2 \ell + 1)!!}{(\ell + \ell' + 1)!!} \sum_{\substack{\vec{m} \\
  0 \le m_i \le \ell'/2 \\ m = (\ell' - \ell)/2}} \frac{1}{2^m} \,
\frac{\ell_x'!}{(\ell_x'-2m_x)! \, m_x!} \,
\frac{\ell_y'!}{(\ell_y'-2m_y)! \, m_y!} \,
\frac{\ell_z'!}{(\ell_z'-2m_z)! \, m_z!} \, \langle \vec{\ell} | \vec{\ell}' - 2 \vec{m} \rangle
  \, ,
\label{eq:lrproduct}
\end{eqnarray}
where in the second equality we have made use of the expansion (\ref{eq:nexpansion})
for the product of direction vector components.

In practical circumstances, the sums over $\ell$ must be cut off at some $\ell_{\rm max}$.
This is because, on one hand,
the experimental apparatus and the statistic limit the
angular resolution and, on the other,
the features of a reaction limit the number of $\ell$ for which $C_{\vec{\ell}}$ may be significant.
Exceeding a sensible $\ell_{\rm max}$ will result in coefficients dominated by noise.  A likely situation
is that of a suitable value of $\ell_{\rm max}$ higher for $X$, $B$ and $\bar{C}$ than for $C$.
This is because the anisotropies associated with measurement efficiencies, particularly at small~$q$,
can be much stronger than physical anisotropies, especially when the experiment is blind in certain directions.  In the
latter case, the binning method would necessarily fail for small bins,
while the method relying on the $\vec{\ell}$-decomposition
from start could succeed when using large $\ell_{\rm max}$ for $X$ and $B$, as long as only low $\ell$-values were
demanded of $C$.

Strategies similar to that employed for finding $C$ in Eq.~(\ref{eq:XBC}) may be employed for finding the
expansion coefficients for an inverse of an angular function or for a product of angular functions.
In finding the inverse
of $B$ in terms of
$B_{\vec{\ell}}$, one would replace $X(\Omega)$ in~(\ref{eq:XBC}) by 1, leading to $X_{\vec{\ell}} = \delta_{\ell 0}$ in
Eqs.~(\ref{eq:C0=}) and ~(\ref{eq:Cl=}).  On the other hand,
in looking for the product of $B$ and $C$ in terms of expansion coefficients,
one can first obtain the coefficients
\begin{equation}\label{eq:Xl=}
  \bar{X}_{\vec{\ell}} =  \frac{1} {\gamma(\vec{\ell})} \, \sum_{\substack{ \vec{\ell}' \\ 0 \le \ell_i' \le \ell_i}}
  \gamma(\vec{\ell}-\vec{\ell}') \,
\gamma(\vec{\ell}') \,
B_{\vec{\ell} - \vec{\ell}'} \, {C}_{\vec{\ell}'} \, ,
\end{equation}
that do not meet the condition of tracelessness.  Next, one can obtain
traceless coefficients ${X}_{\vec{\ell}}$ following such a procedure as for $C$ in
Eqs.~(\ref{eq:Cell})-(\ref{eq:lrproduct}).  Obviously, that specific detracing procedure is a form of application of
the projection operator ${\mathcal P}$, described in the appendix, to a symmetric tensor.

We have, thus, demonstrated some practical procedures for dividing, inverting and multiplying angular functions
in terms of $\vec{\ell}$-arrays without resorting to binning of the functions in
$\cos\theta$ or $\phi$ and without performing matrix inversion operations.

\section{Practical Aspects of Shape Analysis}
\label{sec:source}

The expansion of correlations and sources in tesseral or cartesian harmonics
brings about practical benefits.
The three-dimensional dependencies
on relative momentum and spatial separation of the functions are replaced by sets of one-dimensional
dependencies of angular expansion coefficients.  The replacing of the three-dimensional by one-dimensional
dependencies allows,
principally, for a more thorough representation of the correlation data
and of the physical picture of those data's implications.
The three-dimensional integral relation
between the correlation function and source gets replaced by a set
of one-dimensional relations between the respective expansion coefficients, simplifying the task
of finding the source features.
The expansion coefficients for the functions, as either spherical or cartesian tensors,
transform covariantly under rotations.

Of the two sets of harmonic functions for expansion, the tesseral set represents an orthonormal basis,
but leads to coefficients
which are complex and hard to interpret.  On the other hand, the cartesian set is overcomplete,
generally nonorthogonal within any rank,
but yields real coefficients with relatively straightforward interpretations.  The transformation properties,
under rotations, are more straightforward for the cartesian than for the tesseral coefficients.
We have shown,
in the preceding section, how to transcribe results between the two harmonic representations.

\subsection{Shapes in Terms of Cartesian Harmonics}

Along any direction of relative momentum or separation, the correlation or source function are equal to
combinations of the cartesian coefficients (\ref{eq:cartesianmaster_exp}).  Particularly simple results
follow along any of the employed cartesian axes; e.g.\ along the positive direction
for the $x$-axis the correlation function is:
\begin{equation}\label{eq:Rxaxis}
  {\mathcal R}({q}) = {\mathcal R}^{(0)}({q}) + {\mathcal R}^{(1)}_{x} ({q}) + {\mathcal R}^{(2)}_{xx} ({q})
  + {\mathcal R}^{(3)}_{xxx} ({q})  + \ldots \, .
\end{equation}
Terms for subsequent $\ell$ refine the information on the
dependence of the function on $q$, along the direction.
As we have mentioned, the physics, on one hand, and statistics and apparatus, on the other, will limit
the number of significant terms in the expansion, with the high-$\ell$
terms being dominated by noise \cite{Brown:2005ze}.

For identical particles, the correlation function and accessed emission source are invariant under inversion.
In such a case, the odd-$\ell$ terms in the harmonic expansion vanish.  That is, generally, not the case for
nonidentical particles.  For instance, if the protons in a reaction are, on the average,
emitted earlier than pions, the distribution ${\mathcal S}$ of pairs in ${\pmb r}_p - {\pmb r}_{\pi}$ will
be pushed out in the outward direction, away from~0.  The three $\ell=1$ cartesian moments (\ref{eq:nexpansion}),
computed from~${\mathcal S}^{(1)}_\alpha$,
yield the magnitude and precise direction of the average displacement of
the distribution.  At individual~$r$, the effects onto ${\mathcal S}$ of different emission times
and possibly of different
emission locations will result in a dipole distortion representable in terms of magnitude and
direction:
\begin{equation}\label{eq:S1=}
  {\mathcal S}_\alpha^{(1)}(r) = S^{(1)}(r) \, e_\alpha^{(1)}(r) \, .
\end{equation}
In the above, the direction vector is ${\pmb e}^{(1)}= [\sin{\theta^{(1)}} \cos{\phi^{(1)}},
\sin{\theta^{(1)}} \sin{\phi^{(1)}}, \cos{\theta^{(1)}} ]$.  Not only the magnitude of the angular distortion,
but also the characteristic directions of the distortion can depend on the relative distance.
If the distortion direction is independent of the distance, this direction will be
further the same for the average displacement and the same for the correlation at any relative
momentum.  The latter is due to the fact that the kernel ${\mathcal K}_\ell$ is universal within a given rank,
i.e.\ the same
for different tensor components.
The $\ell=1$ distortions are behind the effects investigated by Gelderloos {\em et al.}\
\cite{gelderloos} and Lednicky {\em et al.}\ \cite{lednicky},
who have compared pair emission probabilities for pairs
with relative momenta directed along and opposite to the pair total momentum.
The~advantage of using the cartesian harmonics is the ability to determine the
distortion directions in correlation, and the implication of those distortions for source distortions,
without any presumption regarding the distortion direction when no possible symmetry arguments can be invoked.

Quadratic moments of the emission source are associated with the rank $\ell=2$ and $\ell=0$ expansion
coefficients, cf.~Eq.~(\ref{eq:cartesiansourcemoments}), with $\ell=2$ rank
giving rise to an anisotropy of the tensor
out of the moments, $\langle r_{\alpha_1} \, r_{\alpha_2} \rangle$.  A~triaxial ellipsoid
could be associated with those moments, with Euler angles
describing rotation from the coordinate axes to the axes of the
ellipsoid, i.e.\ the axes along which the tensor of moments gets diagonalized.
At any value of the corresponding relative variable, for either the source or the correlation function,
the rank $\ell=2$ symmetric and traceless tensor of
expansion coefficients may be diagonalized and
described in terms of 5 independent parameters, 2 distortions and 3 angles:
\begin{equation}\label{eq:S2=}
  {\mathcal S}_{\alpha_1 \, \alpha_2}^{(2)}(r)
  = S_1^{(2)} \, e_{1 \, \alpha_1}^{(2)} \, e_{1 \, \alpha_2}^{(2)}  -
  (S_1^{(2)} + S_3^{(2)}) \, e_{2 \, \alpha_1}^{(2)} \, e_{2 \, \alpha_2}^{(2)}
  + S_3^{(2)} \, e_{3 \, \alpha_1}^{(2)} \, e_{3 \, \alpha_2}^{(2)} \, .
\end{equation}
The three eigenvectors ${\pmb e}_i^{(2)}$ in (\ref{eq:S2=}) form an orthonormal set.  Due to the tracelessness of
${\mathcal S}^{(2)}$, the distortions associated with the three directions add up to zero,
$S_2^{(2)} = - (S_1^{(2)} + S_3^{(2)})$.
In terms of the Euler angles $\Phi^{(2)}$, $\Theta^{(2)}$ and $\Psi^{(2)}$, for rotation
of unit vectors along coordinate axes to
$\lbrace {\pmb e}_i^{(2)} \rbrace_{i=1,2,3}$, the eigenvectors may be represented as
\begin{eqnarray}
{\pmb e}_1 & = & [\cos{\Psi} \, \cos{\Phi} - \cos{\Theta}  \, \sin{\Psi} \,  \sin{\Phi},
\cos{\Psi} \,  \sin{\Phi} + \cos{\Theta} \,  \cos{\Psi} \,  \sin{\Phi}, \sin{\Psi} \,  \sin{\Theta}] \,
, \nonumber \\
{\pmb e}_2 & = & [-\sin{\Psi} \,  \cos{\Phi} - \cos{\Theta} \,  \sin{\Psi} \,  \cos{\Phi},
-\sin{\Psi} \,  \sin{\Phi} + \cos{\Theta} \,  \cos{\Psi} \,  \cos{\Phi}, \cos{\Psi} \,  \sin{\Theta}] \,
, \nonumber \\
{\pmb e}_3 & = & [\sin{\Theta} \, \sin{\Phi}, - \sin{\Theta} \, \cos{\Phi}, \cos{\Theta}] \, ,
\end{eqnarray}
where the $(2)$ superscripts are suppressed.  Just as the distortion values,
the angles may depend on $r$.  If the angles do not depend on $r$, then
those angles coincide with those for the diagonalization of the tensor of quadratic moments,
$\langle r_{\alpha_1} \, r_{\alpha_2} \rangle$, and for the diagonalization of
the quadrupole coefficients of the
correlation function at any~$q$.  The 5 independent parameters at any relative variable correspond
to 5 independent expansion coefficients for rank $\ell=2$.

With regard to the higher, $\ell > 2$, shape distortions, one way to associate the directions of axes
with them, is to choose one direction, defined by ${\pmb e}_1^{(\ell)}$, by demanding that
the distortion along that direction,
equal to
\begin{equation}\label{eq:SOmega}
 {S}^{(\ell )}(\Omega)
 =\sum_{\alpha_1 \ldots \alpha_\ell} {\mathcal S}_{\alpha_1 \ldots \alpha_\ell}^{(\ell)} \,
e_{1 \, \alpha_1}^{(\ell)} \ldots e_{1 \, \alpha_\ell}^{(\ell)} \, ,
\end{equation}
is
extremal, minimal or maximal \cite{Danielewicz:2005qh}.
Directing a coordinate axis along ${\pmb e}_1^{(\ell)}$, makes the two
${\mathcal S}^{(\ell)}$-coefficients, characterized by $\ell_1 = \ell -1$, vanish.
The next axis, defined by ${\pmb e}_2^{(\ell)}$, may be chosen again as one
extremizing the distortion, but now within the direction plane perpendicular to~${\pmb e}_1^{(\ell)}$.
When directing the second coordinate axis along ${\pmb e}_2^{(\ell)}$
and the third along the direction perpendicular
to the first two, we find that the choice puts to zero the
${\mathcal S}^{(\ell)}$-coefficient characterized by $\ell_1 =0$,
$\ell_2 = \ell - 1$ and $\ell_3 = 1$.

\subsection{Harmonic Decomposition vs Fits in Three-Dimensions}
\label{section:advantages}

For a given total momentum, the~correlation function contains three-dimensional information
pertinent to the three-dimensional structure of
the source function.  Effectively, there are three competing approaches for handling the information
in the correlation function.  In the first approach, most common till now in literature,
the data is presented along chosen discrete
directions of relative momentum, typically one or three.  In the second approach the data is fitted to a parameterized form, e.g., a three-dimensional Gaussian, in the full three-dimensional space of relative momentum.
When going beyond the simplest parameterizations in this
approach, it may be difficult to assess, working in the
three-dimensional space, which aspects of the source parameterization
affect which particular aspects of
the correlation.
In the third approach, the data is expanded in harmonic functions.  A
set of one-dimensional functions then yields a complete representation of the three-dimensional
data.  Upon decomposition, the deduction of source features is reduced to a set of one-dimensional
problems, with one-to-one correspondence between the coefficients of the correlation and
the source.  A specific coefficient of the source affects the specific coefficient of the correlation
function and no other independent coefficients.  Whether one were to advance the source analysis
as an inversion problem, or as a problem of fitting, it should be certainly preferable
to divide and conquer, i.e.\ approach the issue as a set one-dimensional problems.

Spherical harmonic decompositions generally carry more information than
can be expressed with parameters of gaussian sources employed in the literature.
The integrated
nine independent source
coefficients, ${\mathcal S}^{(\ell)}$, for $\ell \le 2$, fully determine the moments
$\langle r_{\alpha_1} \, r_{\alpha_2} \rangle$ and $\langle r_\alpha \rangle$, for all
$\alpha$-indices, cf.\ Eq.~(\ref{eq:nexpansion}).
Those moments and the integral of~$S^{(0)}$, $\lambda = 4 \pi
\int {\rm d}r \, r^2 \, {\mathcal S}^{(0)}(r)$, determine all the ten parameters that can be set for
a gaussian source such as discussed in~\cite{Danielewicz:2005qh} or in the next section.
However, the coefficients ${\mathcal S}^{(\ell)}$, for $\ell \le 2$, may have a different
dependence on $r$ than those for a gaussian source.  In addition, the $\ell > 2$ coefficients
may further yield different characteristic moments.  When coordinate
axes are directed along the principal axes of the gaussian quadrupole deformation,
the gaussian moments, relative to gaussian center, factorize:
\begin{equation}
\label{eq:factorization}
\left \langle (r_1-\langle r_1 \rangle)^{\ell_1} (r_2-\langle r_2 \rangle)^{\ell_2}
(r_3 - \langle r_3 \rangle)^{\ell_3} \right \rangle=
\left \langle (r_1-\langle r_1 \rangle)^{\ell_1} \right \rangle
\left \langle (r_2-\langle r_2 \rangle)^{\ell_2} \right \rangle
\left \langle (r_3-\langle r_3 \rangle)^{\ell_3} \right \rangle \, .
\end{equation}
For distinguishable particles in the midrapidity region of symmetric ulrarelativistic collisions, 
the above
implies that $\langle r_{\rm out} \,
r_{\rm long}^2\rangle = \langle r_{\rm out} \rangle \langle
r_{\rm long}^2 \rangle $.  The latter factorization will be violated for instance for boost-invariant
particle emission, such as discussed in Sec.~\ref{subsec:example_blast}.  In the case
of boost-invariant emission, with a~transverse expansion,
the single-particle sources acquire a boomerang-type shape, as particles
at lower positions along the beam axis get emitted locally earlier than particles at higher
positions.  The violation of the factorization above will become
apparent if one e.g.\ observes that the boomerang shape will be preserved in the relative
source when one of the emitted particles is heavy (baryon) and the other light ($\pi$~meson).
Features of a boomerang shape may be quantified with the source expansion coefficients for
$\ell=3$, which contribute to the moment $\langle r_{\rm out} \,
r_{\rm long}^2\rangle$, cf.~(\ref{eq:nexpansion}), that may be compared to the product
$\langle r_{\rm out} \rangle \langle
r_{\rm long}^2 \rangle $ from $\ell=1$ and $\ell=2$, besides~$\ell=0$.

With regard to the dependence of ${\mathcal S}^{(\ell)}$ on $r$, already several studies have
focused on determining the non-gaussian behavior of the angle-averaged source function,
${\mathcal S}^{(\ell=0)}(r)$.  With regard to $\ell \ge 1$, significant values of
${\mathcal S}^{(\ell)}(r)$ are principally expected only there where
${\mathcal S}^{(\ell=0)}(r)$ begins
to fall, i.e.\
around the edges of the source, whereas, at small~$r$, ${\mathcal S}^{(\ell)}(r) \propto r^\ell$.
Concerning the low-$r$ region, in the source-correlation relation, the convolution with
kernel ${\mathcal K}_\ell$ brings in a factor $r^2$.  That low-$r$ suppression generally
reduces the impact of
small-$r$ $S^{(\ell)}$
on the correlation, enhancing errors in the restored source, making it likely difficult to detect
any nongaussian features there at $\ell \ge 1$.  On the other hand, in the source tails
at $\ell \ge 1$ in the
source coefficients, the non-gaussian behavior may
be actually enhanced compared to $\ell=0$, both at low and high reaction energies.  
For instance, long-lived slowly cooling systems
with emission rates falling off
as ${\rm e}^{-t/\tau}$ should give rise to relative sources with an exponential fall-off along
the direction that the pair total momentum has in the frame of the emitting system.  For low $\ell$,
strength of the exponential tail should increase with~$\ell$ for distortion along the direction
of the total momentum, Eq.~(\ref{eq:SOmega}).  In fact, looking for the maximal distortion in source
tail may help to locate the emitting source within the velocity space.  In ultrarelativistic
collisions, the emission becomes boost-invariant around the midrapidity.  For a freezing-out
boost-invariant thermal source, the single-particle distribution of emission points in the
space-time, cf.\ Eq.~(\ref{eq:Sss}), has the form
\begin{equation}
\label{eq:dNther}
s({\pmb p},{\pmb r},t) \sim \delta(\tau-\tau_f) \, {\rm e}^{-E_\perp \sqrt{1+z^2/\tau^2}/T} \, ,
\end{equation}
where $\tau = \sqrt{t^2-z^2}$ and $z$ points along the beam axis.  The single-particle source above
falls off exponentially with an increase in $|z|$ and so would the relative source with an increase
in $|r_{\rm long}|$.  This behavior should be particularly revealed in the ${\mathcal S}_{[0,0,2k]}$
coefficients of the relative source.

While it may be straightforward to compute the expansion coefficients as convolutions of the
correlation with different harmonics, some caution should be exerted in selecting
the range of relative momentum $q$ for the source investigation.  The caution is required due to the fact
that there exist other causes for inter-particle correlation than
the final-state effects within a two-particle system.  At the mathematical level, those
causes can give rise to a ${\pmb q}$-dependence of the source ${\mathcal S}$, assumed to be negligible
in arriving at Eq.~(\ref{eq:corrmaster}).  The~dependence might be produced by a collective
expansion of strength comparable or larger than the local thermal motion in the emitting region.
If the dependence is not explicitly corrected for, the analysis for source restoration should
be limited to the region where the source ${\pmb q}$-dependence is deemed weak.
In case of the final-state effects due to resonances,
short-distances in ${\mathcal S}$ are of importance, with the $q$-dependence governed there by local
temperature at freeze-out, giving rise to the requirement of $q \ll \sqrt{2 \mu \, T}$ for the source analysis.  This
condition is practically always satisfied for $pp$ pairs whose correlation function
peaks at $q \sim 20$~MeV/c.  On the other hand, for $p \pi^+$ pairs in the region of $\Delta$-resonance
at $q \sim 225$ MeV/c, the condition is practically never satisfied in nuclear reactions.
Correlations due to directed flow associated with the reaction plane, such as due to elliptic 
and sideward flows,
can be partially accounted for by controlling the reaction plane in constructing
the denominator in~(\ref{eq:corrmaster}).  Otherwise, those type of correlations tend to peak at $q$ of
the order of the mean $p_\perp$ (i.e.\ a few hundred MeV/c for $p$ or $\pi$ in energetic collisions), 
reaching there
values of the order of few percent.  By comparison, Coulomb correlations fall off with increase of $q$, as
$1/q^2$, down to about one percent for $pp$ typically at $q \sim 100$~MeV/c.  That further underscores the need to
limit the range of $q$ in source analysis.  In ultrarelativistic collisions, jets are an additional cause
for inter-particle correlations.  The characteristic scale for those correlations is of the order of several
hundred MeV/c and the magnitude of the order of one percent towards lower~$p_\perp$.  Overall, it is apparent
that caution should be executed when considering small values of the ${\mathcal R}$, i.e.\ of the order of half a percent or less.  If conclusions are to hinge on such small values, some supporting evidence
should be provided to justify the assumption that the correlation originates from final-state interactions or symmetrization.

\section{Examples of Emission}
\label{sec:examples}

Here, we provide examples of what may be expected in reactions regarding harmonic
characterization of emission sources and of correlation functions.

\subsection{Gaussian Sources}
\label{subsec:example_gaussian}

In the analyses of correlation functions, it is quite common to aim at a gaussian
representation of the emission sources.  When cast in the gaussian form, a source
is expressed in terms of a minimal number of parameters needed for shape description.

Thus, a gaussian source is generally described in terms of 9 parameters, out which 3 describe
the position of the source center ${\pmb d}$ and 6 correspond to the independent
independent matrix elements of the symmetric positive shape matrix~$M$:
\begin{equation}
\label{eq:Sgauss}
{\mathcal S}({\pmb r}) = \frac{\left[ \det{M} \right]^{1/2} }{(4 \pi)^{3/2}}
\exp{ \left\lbrace
- \frac{1}{4} \sum_{\alpha_1 \, \alpha_2} M_{\alpha_1 \, \alpha_2} \,
(r_{\alpha_1} - d_{\alpha_1}) \, (r_{\alpha_2} - d_{\alpha_2})
\right\rbrace
 } \, .
\end{equation}
A tenth parameter would describe the normalization of
the source, which could vary from unity if a significant fraction of the pairs had
effectively infinite separations due to weak decays and were outside the description in terms of
a single gaussian. For identical particles, the accessed relative source must be symmetric under
inversion, representing particle interchange.
In that case ${\pmb d}=0$, leaving only 6 independent shape parameters for the source.

The convenient choice for the shape
parameters are three Euler angles describing an orientation
of the eigenvectors~$\lbrace {\pmb u}_i \rbrace$ of $M$ and
three gaussian radii $R_i$ in terms of which the matrix
is $M_{\alpha_1 \, \alpha_2} = \sum_i (1/R_i^2) \, {\pmb u}_{i \, \alpha_1} \, {\pmb u}_{i \, \alpha_2}$.
The specific factor of $(1/4)$ multiplying $M$ and $(1/R^2)$ in (\ref{eq:Sgauss}) is associated with the
tradition of assigning the radii to the single-particle sources $s$ in (\ref{eq:Sss}).
However, absolute positions in a reaction are not observable;
for the description of relative positions more appropriate
would be a factor of $(1/2)$ yielding radii larger by a factor of $\sqrt{2}$.  When coordinate axes
are oriented along the gaussian shape eigenvectors, the gaussian shape takes the form:
\begin{equation}
{\mathcal S}({\pmb r}) = \frac{1}{(4 \pi)^{3/2} \, R_x \, R_y \, R_z}
\exp{ \left\lbrace
- \frac{(x - d_x)^2}{4 R_x^2} - \frac{(y - d_y)^2}{4 R_y^2}
- \frac{(z - d_z)^2}{4 R_z^2}
\right\rbrace
 } \, .
\end{equation}

The displacement ${\pmb d}$ of the source for distinguishable particles, relative to ${\pmb r} = 0$, is
of interest in both intermediate-energy
and high-energy collisions \cite{lednicky,panitkinoffset}.
For instance, at low energy it might be of interest whether neutron emission
preceded proton emission \cite{Ghetti:2004ac}, producing a ${\pmb d}$ in the
direction of pair total momentum, or whether intermediate-mass
fragments were emitted simultaneously with the protons \cite{gelderloos}.  At high energy, one might
look for evidence of strange particles leaving early
\cite{Wang:1999bf}.  A simultaneous emission combined with radial
expansion should lead to heavy particle emission shifted out in the direction of
expansion relative to light particle emission \cite{Retiere:2003kf}.

In many circumstances, the structure of anisotropies for measured correlation functions and corresponding
structure for imaged sources simplifies.
Thus, if the reaction plane is not identified, the measured
correlation function and the corresponding source must obey reflection symmetry with respect to the plane
formed by the pair total momentum ${\pmb P}$ and the beam axis.  In the gaussian representation,
the displacement vector ${\pmb d}$ must lie within that plane and so must two of the eigenvectors of the gaussian shape
matrix~$M$.  In the midrapidity region of a symmetric system, the correlation function and source must be
moreover
symmetric with respect to forward-backward reflection.  That makes the ${\pmb d}$-vector perpendicular to
the beam axis and makes the in-plane gaussian-shape eigenvectors pointing along and perpendicular to the
beam axis \cite{Wiedemann:1999qn}.

The nine gaussian parameters discussed above may be related both to the angular moments of the source
and of the correlation function.  It may be tempting, in particular,
to associate the three independent $\ell=1$ moments with the
displacement ${\pmb d}$ and the five $\ell=2$ moments with the shape anisotropy of a gaussian.
However, the situation is more involved.  For illustration, one can consider an isotropic gaussian
displaced in the $x$-direction:
\begin{eqnarray}
{\mathcal S}({\pmb r}) & = & \frac{1}{(4 \pi)^{3/2} \, R^3}
\exp{ \left\lbrace
- \frac{(x - d)^2 + y^2 + z^2}{4 R^2}
\right\rbrace
 } = \frac{1}{(4 \pi)^{3/2} \, R^3}
\exp{ \left\lbrace
- \frac{r^2 + d^2}{4 R^2}
\right\rbrace} \exp{ \left\lbrace \frac{x d}{2 R^2} \right\rbrace}
\nonumber \\
&= &  \frac{1}{(4 \pi)^{3/2} \, R^3}
\exp{ \left\lbrace
- \frac{r^2 + d^2}{4 R^2}
\right\rbrace}
\sum_{\ell=0}^\infty (2 \ell+1) \, I_\ell \left( \frac{r d}{2 R^2} \right) \, P_\ell \left( n_x \right)
\nonumber \\
&= &  \frac{1}{(4 \pi)^{3/2} \, R^3}
\exp{ \left\lbrace
- \frac{r^2 + d^2}{4 R^2}
\right\rbrace}
\sum_{\ell=0}^\infty \frac{(2 \ell+1)!!}{\ell!} \, I_\ell \left( \frac{r d}{2 R^2} \right) \,
{\mathcal A}_{xx \ldots x}^{(\ell)} (\Omega)
 \, .
\label{eq:besselmess}
\end{eqnarray}
Here, $I_\ell$ are the modified spherical Bessel functions of the first kind.  For small argument values
of those functions,
the functions are proportional to the power of the argument, $I_\ell(x) \propto x^\ell$.
Equation (\ref{eq:besselmess}) illustrates that the source displacement ${\pmb d}$ produces non-zero angular
moments for all $\ell \ge 1$ even when the gaussian source is isotropic around its center.
Incidentally, in a similar manner,
for ${\pmb d}=0$, the anisotropy of a gaussian gives rise to non-zero angular moments
for all even $\ell \ge 2$,
not just $\ell=2$.

For a general displaced and anisotropic gaussian, the angular coefficients can be obtained, as a function
of $r$, through a direct angular integration.  Otherwise, the power series for an exponential may be employed,
in combination
with the properties of spherical harmonics, to yield converging series for the coefficients.
Specifically, as the argument of a source gaussian
is quadratic in coordinates, that argument can be represented in terms of
harmonics of rank $\ell \le 2$, leading to:
\begin{equation}
\label{eq:GammaExp1}
{\mathcal S} ({\pmb r}) = c (r) \, {\rm e}^{\Gamma ({\pmb r})} \, ,
\end{equation}
where
\begin{equation}
\label{eq:GammaExp2}
\Gamma ({\pmb r}) = \sqrt{4 \pi}
\sum_{\substack{ 1 \le \ell \le 2  \\ m }} \Gamma_{\ell m}^*(r) \,
Y_{\ell m} (\Omega) = \sum_{\substack{ \vec{\ell} \\ 1 \le \ell \le 2}} \gamma(\vec{\ell}) \, \Gamma_{\vec{\ell}} (r) \,
\hat{n}^{\ell_x} \, \hat{n}^{\ell_y} \, \hat{n}^{\ell_z} \, .
\end{equation}
The exponential ${\rm e}^{\Gamma}$ may be next expanded in the power series yielding,
\begin{equation}\label{eq:expGamma}
 {\mathcal S} ({\pmb r}) =
 \sum_{N=0}^{\infty} {\mathcal S}_N ({\pmb r}) \, ,
\end{equation}
where
\begin{equation}\label{eq:GammaN}
  {\mathcal S}_N ({\pmb r}) = {\mathcal S}_{N-1} ({\pmb r}) \, \frac{\Gamma ({\pmb r})}{N} \, ,
\end{equation}
and ${\mathcal S}_0 (r) = c(r)$.
Individual terms of the sum in (\ref{eq:expGamma}) may be represented in terms of the surface spherical
harmonics of either type, leading to corresponding expansion for the source~${\mathcal S}$.
The rules of superposition for tesseral harmonics yield the following recursion relation for the
tesseral expansion coefficients:
\begin{equation}\label{eq:GNrecursion1}
  {\mathcal S}_{N \ell m} (r) = \frac{\sqrt{4 \pi}}{N} \sum_{\ell' \, \ell'' \, m'}
  (\ell' \, m' \, \ell'' \, (m-m') | \ell \, m) \, \Gamma_{\ell' \, m'} (r) \,
  {\mathcal S}_{(N-1) \, \ell'' \, (m-m')} (r) \, ,
\end{equation}
where $(\cdot | \cdot)$ represents a Clebsch-Gordan coefficient.  It follows that
terms in the sum (\ref{eq:expGamma})
decrease rapidly with $N$, for $N \gg \max{}_{\ell m} |\Gamma_{\ell m} (r)|$.  When aiming at the
source in terms of
cartesian harmonic coefficients, first the cartesian coefficients $\bar{\mathcal S}_{N \vec{\ell}}$
which do not meet the tracelessness
condition are found through recursion:
\begin{equation}\label{eq:GNrecursion2}
  \bar{\mathcal S}_{N \vec{\ell}}(r) = \frac{1}{N \, \gamma(\vec{\ell})} \sum_{\vec{m}}
  \gamma(\vec{m}) \, \gamma(\vec{\ell} - \vec{m}) \, \Gamma_{\vec{m}}(r) \,
  \bar{\mathcal S}_{(N-1) \, (\vec{\ell} - \vec{m})}(r) \, .
\end{equation}
In the next stage, the traceless coefficients ${\mathcal S}_{N \vec{\ell}}$ are found
following such a procedure as
in
Eqs.~(\ref{eq:Cell})-(\ref{eq:lrproduct}).  Finally, the coefficients for the source may be obtained by
summing over the respective series:
\begin{equation}\label{eq:SNlm}
  {\mathcal S}_{\ell m}(r) = \sum_{N=0}^{\infty} {\mathcal S}_{N \ell m} (r) \, ,
  \hspace*{2em}\mbox{or} \hspace*{2em}
  {\mathcal S}_{\vec{\ell}}(r) =  \sum_{N=0}^{\infty} {\mathcal S}_{N \vec{\ell}} (r) \, ,
\end{equation}
depending whether the tesseral or cartesian harmonics are followed.  Coefficients
for gaussian sources in this Section have been obtained following the recursion outlined above.
Kernels for use in Eqs.\ (\ref{eq:ylmmaster}) and (\ref{eq:cartesianmaster}) have been
obtained using~\cite{Pratt:2003ar} known phase-shifts.

\begin{figure}
\centerline{\includegraphics[width=0.4\textwidth]{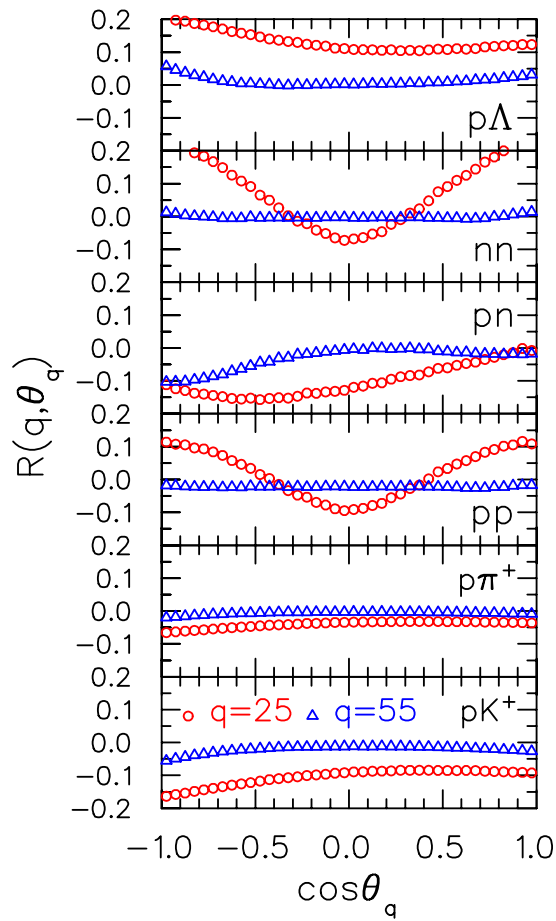}}
\caption{\label{fig:c_vs_ctheta} (Color online)
Correlations associated with a relative source
in the form
of an axially symmetric gaussian, characterized
by $R_x = R_y = 4 \, \mbox{fm}$ and $R_z=8 \, \mbox{fm}$, for different indicated particle pairs.
For distinguishable particle pairs,
the center of this gaussian is shifted by ${\pmb d} = (0,0,4\, \mbox{fm})$ relative to the origin at
$r=0$.
The correlations are shown as a function of cosine of the angle
that the relative momentum ${\pmb q}$ makes with
the $z$-axis, for two values of momentum magnitude, $q=25 \, \mbox{MeV}$ and  $q=55 \, \mbox{MeV}$.
For the specific employed source,
angular anisotropies in the correlation functions are of the same order as the functions. }
\end{figure}

In order to illustrate the possibility of measuring source shapes, we consider a relative
source represented by
an axially symmetric gaussian oriented along the beam axis
with parameters $R_x = R_y \equiv R_\perp = 4 \, \mbox{fm}$ and $R_z=8 \, \mbox{fm}$.  For distinguishable
particles, we additionally displace this gaussian away from the origin at $r=0$ by
${\pmb d} = (0,0,4\, \mbox{fm})$.
Such a source may result in
consequence of emission of the individual species within two single-particle sources
differently elongated along
the $z$ axis, with the source for the $a$ species displaced by $d$ along the $z$-axis relative
to the source for the $b$ species.
For the chosen relative source, the anisotropies of associated correlation functions can be quite
significant.  Examples of the associated correlations, for different particle pairs, $pK^+$,
$p\pi^+$, $pp$, $pn$, $nn$ and $p\Lambda$, are presented in Fig.~\ref{fig:c_vs_ctheta}.
The correlations are
shown there as
a function of $\cos{\theta_{\pmb q}}$ at fixed $q$.  The observed angular dependence of
the functions is exclusively due to the source anisotropy.
While the correlations are stronger at low than at high~$q$, the correlations and their
anisotropies are generally of the same order.
In the dependence of correlations on $\cos{\theta_{\pmb q}}$, for distinguishable
particles, one recognize both components odd and even in $\cos{\theta_{\pmb q}}$.  The odd
components are associated with the displacement ${\pmb d}$ in the source while even are
associated with both the displacement ${\pmb d}$ and $R_\perp \ne R_z$.  In the correlations of
identical particles, only the effects of $R_\perp \ne R_z$ are present.

\begin{figure}
\centerline{\includegraphics[width=0.7\textwidth]{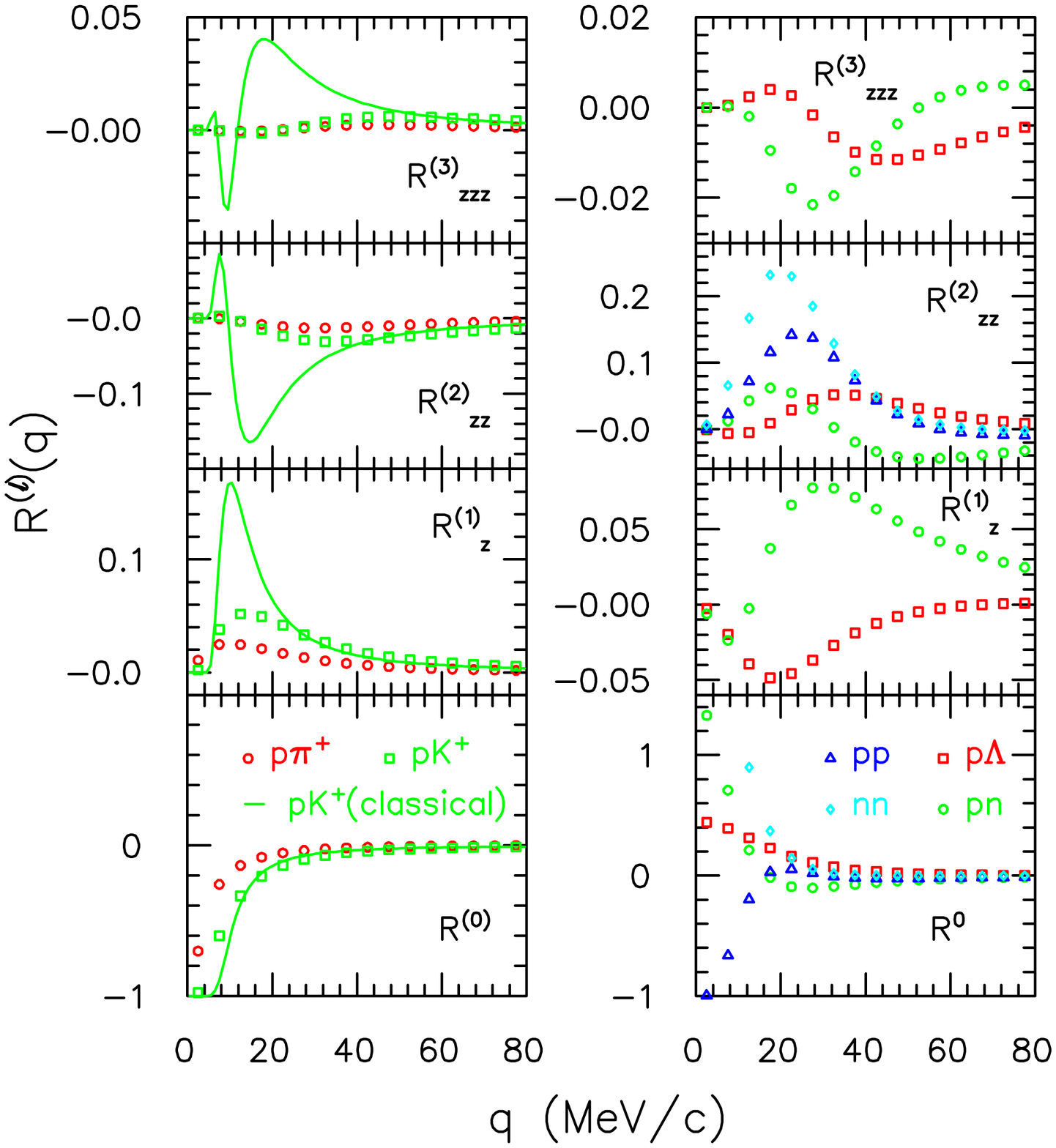}}
\caption{\label{fig:c_gauss} (Color online)
Angular coefficients of the correlations associated with an axially symmetric gaussian
source characterized by $R_x = R_y = 4 \, \mbox{fm}$ and $R_z=8 \, \mbox{fm}$,
for proton-charged meson pairs (symbols in the left panels) and
baryon-baryon pairs (symbols in the right panels),
as a function of the relative momentum~$q$.  For distinguishable particle pairs,
the center of the gaussian source is shifted by ${\pmb d} = (0,0,4\, \mbox{fm})$
relative to the origin at $r=0$.
The $pK^+$ and $p\pi^+$ correlations in the left panels
tend to be dominated by the Coulomb interaction.  To illustrate the degree to which quantal effects
wash out the information on emission source, the left panels show also (solid lines) the $pK^+$
correlations in the limit of classical Coulomb interactions.
}
\end{figure}

With the choice of axial symmetry for the specific source, all cartesian components with odd number
or $x$ or $y$ indices vanish.  For identical particles, this further implies vanishing of
cartesian components with odd number of $z$ indices, since there are no indices left to pair them
with to yield a net even index number.  For distinguishable particles, at $\ell \le 3$, the only nonvanishing
moments are: ${\mathcal S}^{(0)}$,
${\mathcal S}^{(1)}_z$, ${\mathcal S}^{(2)}_{xx}$, ${\mathcal S}^{(2)}_{yy}$, ${\mathcal
S}^{(2)}_{zz}$, ${\mathcal S}^{(3)}_{xxz}$, ${\mathcal S}^{(3)}_{yyz}$ and ${\mathcal
S}^{(3)}_{zzz}$.  Not all of those moments are independent, as ${\mathcal S}^{(2)}_{xx} =
- \left( {\mathcal S}^{(2)}_{yy} + {\mathcal S}^{(2)}_{zz}  \right)$ and
${\mathcal S}^{(3)}_{xxz} = - \left( {\mathcal S}^{(3)}_{yyz} + {\mathcal
S}^{(3)}_{zzz} \right) $, in consequence of the tensor tracelessness.  In addition,
the axial symmetry implies equality between components:
$ {\mathcal S}^{(2)}_{xx} = {\mathcal S}^{(2)}_{yy}$ and
$ {\mathcal S}^{(3)}_{xxz} = {\mathcal S}^{(3)}_{yyz}$.  The left panels of
Fig.~\ref{fig:c_gauss} show the $\ell \le 3$ cartesian coefficients of correlations for
proton-charged meson
pairs.  Those coefficients which are not shown are easily generated from those which are,
as e.g.\ $ {\mathcal S}^{(2)}_{xx} = -{\mathcal S}^{(2)}_{zz}/2  $, etc.

The correlations of protons with charged mesons in Fig.~\ref{fig:c_gauss} are plotted
for $q < 80 \, \mbox{MeV/c}$.  Within that momentum range, the correlation should be
dominated by Coulomb effects.  For comparison purposes, also shown in the figure are
the $p$-$K^+$ correlations expected in the classical limit of the Coulomb interactions.
As expected following the discussions in Sec.\ \ref{sec:kernels}, the quantum
effects seriously affect the correlations at $q R \lesssim \ell + \frac{1}{2}$,
dampening out in that region the classical structures and hampering thus practical
extraction of source shape
information.  With a smaller reduced mass in the $p\pi^+$ than
in the $pK^+$ system, the classical Coulomb hole has a smaller $q$-size in
the $p\pi^+$ system.  In consequence, the quantal effects have a stronger impact on the
features of the $p\pi^+$ correlations, making the extraction of source shape harder from the $p\pi^+$ than
from the $pK^+$ correlations.

\begin{figure}
\centerline{\includegraphics[width=0.4\textwidth]{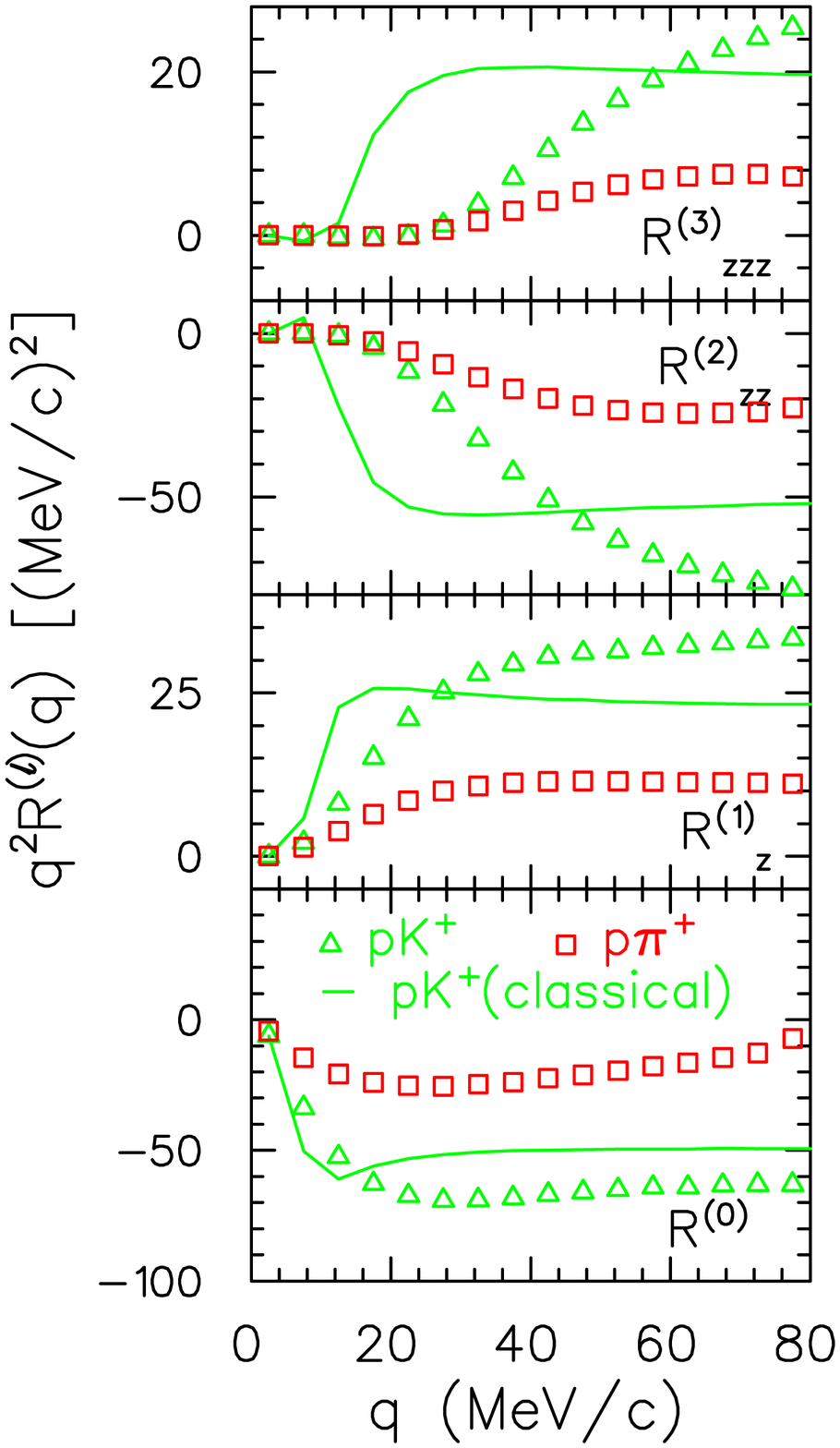}}
\caption{\label{fig:cq2_gauss_mesonN} (Color online) The same proton-charged meson correlation coefficients
as in Fig.~\ref{fig:c_gauss} but now multiplied by the factor of
$q^2$ to better exhibit the features of the correlations at large~$q$.}
\end{figure}

Although difficult, it is certainly tenable to measure correlations of the
order of one percent or less.  In order to exhibit better the correlations at
large $q$, Fig.~\ref{fig:cq2_gauss_mesonN} displays
the same proton-charged meson correlations as those shown in
Fig.~\ref{fig:c_gauss}, but now weighted by $q^2$.  It is apparent that the
correlations decrease as $\ell$ increases.  An important feature of the exhibited correlations is that
they are smooth in $q$ for all $\ell$.  Due to this smooth behavior, it is possible to
analyze those correlations utilizing wide momentum bins.  The phase space
increasing as $q^2$ should further help to garner statistics needed for
analyzing the correlations at the level of one percent or less.  The practical limitation
for the analysis at large $q$ may turn out to be, though, the need to separate
the correlations due to final-state effects within the pair from the possible competing correlations
induced by collective flow and, at RHIC, by jets.

Correlations of elementary baryons are good candidates for source analyses, due
to the large scattering lengths within the pairs which lead to strong correlation structures on
the relative-momentum scale of $q \sim 20\ \mbox{MeV/c}$.  The right-side panels of Fig.~\ref{fig:c_gauss}
display the correlations for $pn$, $nn$, $pp$, $p\Lambda$ pairs, associated with the gaussian
sources.  When present,
the $\ell \ge 1$ baryon-baryon correlations tend to be significantly stronger
at most momenta of experimental interest
than the corresponding Coulomb-dominated meson-baryon correlations.
The odd correlation moments vanish for the $pp$ and $nn$ pairs
as the pair sources and correlations are symmetric under particle interchange.  The
odd-$\ell$ correlations are especially important for the $pn$ pairs
emitted from intermediate-energy heavy ion collisions as these correlations tie to the
issue of whether neutrons are emitted earlier than protons and, further, to the symmetry energy.
Within
high-energy
collisions, odd-$\ell$ $p\Lambda$ correlations are of interest as they can provide information on
whether the hyperons leave earlier than protons.

\subsection{Boost-Invariant Blast-Wave Sources}
\label{subsec:example_blast}

Results of early
pion-pion correlation measurements at RHIC have been satisfactorily
described with fits within the blast wave
models~\cite{Tomasik:2004bn,Retiere:2003kf}.  Those fits have suggested
a nearly instantaneous emission directly contradicting the formation of
a long-lived mixed phase which would have resulted in a dilatory emission and
in relative sources with sizes extended in the outward direction of the pair total
momentum.
Given the
importance of those measurements and findings, it can be important to verify the
conclusions on the nature of particle emission at RHIC by
extracting sources from the correlations of other pairs such as involving kaons or protons.

The minimal blast-wave model used here for illustration employs four parameters
describing the single-particle sources that are boost-invariant along the beam axis:
\begin{equation}\label{eq:1source}
  s({\pmb p},{\pmb r},t) \propto \delta(\tau - \tau_f) \, \theta(r_{\rm max} - r_\perp)
  \, m_\perp \cosh{(y - \eta)} \exp{\left( - \frac{p^\mu \, u_\mu}{T}  \right)} \, .
\end{equation}
Here, $y$ and $\eta$ are, respectively, energy-momentum and space-time longitudinal rapidities.
The emission is assumed to occur at one proper time $\tau_f =10 \, \mbox{fm/c}$.  The decoupling is assumed
to be
spread out over the cylindrical volume of radius $r_{max}=13 \, \mbox{fm}$ and take place in the presence
of a collective motion characterized by transverse rapidity~$\rho$ ($u_\perp = \sin{\rho}$)
proportional to transverse position, $\rho  \propto r_\perp$, producing the maximal
transverse velocity in the longitudinal center of mass of $v_\perp^{\rm max} = \tanh{ \rho^{\rm max} }
= 0.75 \, \mbox{c}$, at $r_\perp = r_{max} $.
The breakup temperature is assumed to be $T = 105 \, \mbox{MeV/c}$.
The chosen parameters describe~\cite{Retiere:2003kf} particle
spectra at $p_\perp < 1 \, \mbox{GeV/c}$, as well as
basic aspects of pion-pion correlations, in Au + Au collisions at
$\sqrt{s_{NN}}= 130 \, \mbox{GeV}$.  Here, our goal is not to fit data,
but to use the blast-wave model to see what results might be expected for
correlations of other particle pairs analyzed in terms of ${\mathcal
R}^{(\ell)}(q)$.

\begin{figure}
\centerline{\includegraphics[width=0.7\textwidth]{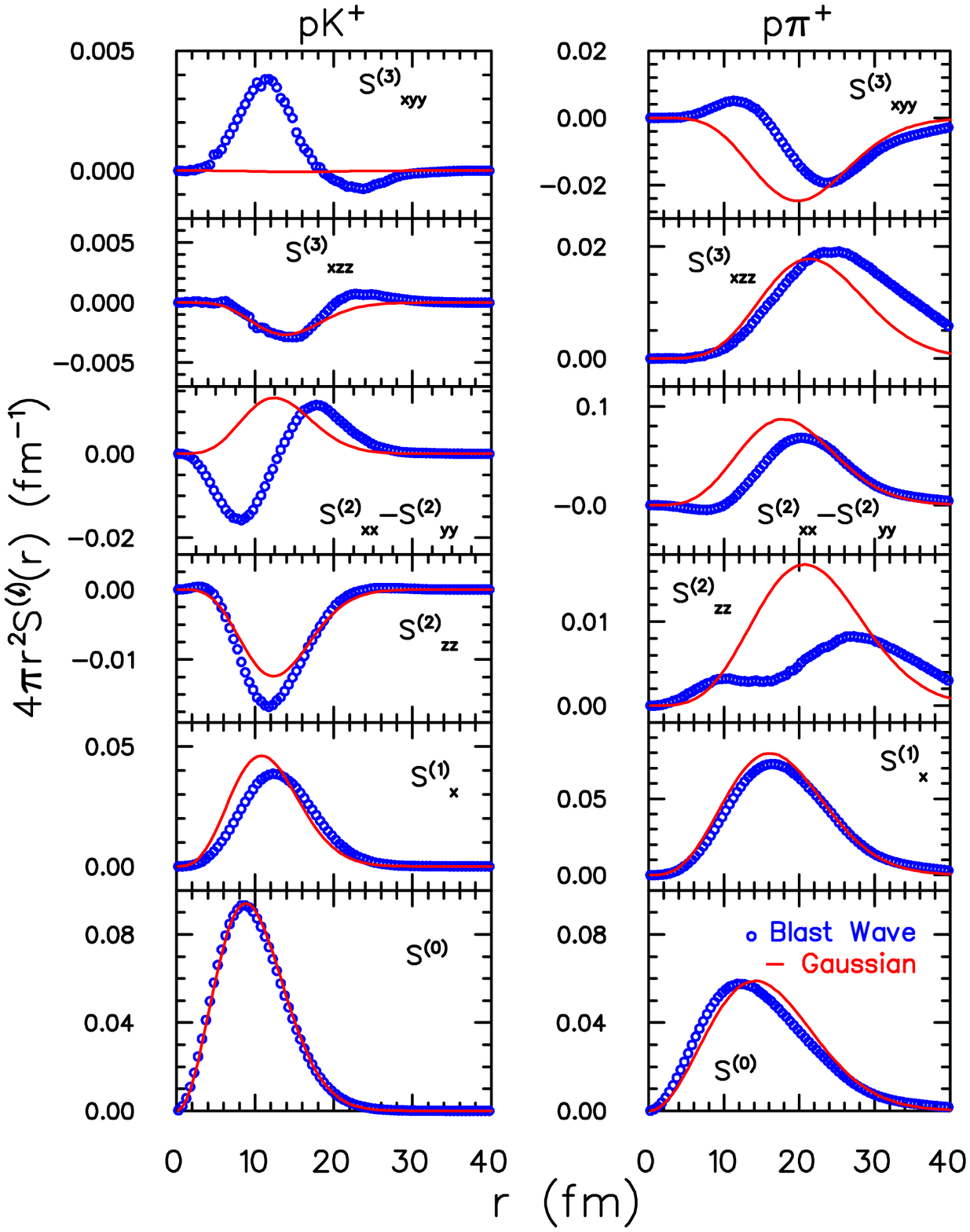}}
\caption{\label{fig:sr2_blast} (Color online)
Angular cartesian coefficients, at $\ell \le 3$, multiplied by $4\pi \, r^2$,
of relative sources for $pK$ (left) and
$p\pi$ (right) pairs moving out at transverse velocity of $0.6 \, \mbox{c}$ relative to the beam axis.
Symbols represent the results obtained in the blast-wave model.  Lines represent the results from
a gaussian source with the same pair c.m.\ cartesian moments, $\langle r_i\rangle$ and
$\langle r_i \, r_j\rangle$, as the blast-wave source.  The non-gaussian features of
a blast-wave source become more apparent as the mass difference within a pair increases.
}
\end{figure}

We focus on correlations of nonidentical particles, because of the possibility
for associated
finite odd angular moments.  The $\ell \le 3$
relative-source moments
for
the $pK$ and $p\pi$ pairs moving at transverse velocity of $v_\perp = 0.6 \,
\mbox{c}$ are shown in Fig.~\ref{fig:sr2_blast}.  To obtain the displayed
results,
emission positions ${\pmb r}_{a,b}$ for each of the
particles $a$ and $b$ had been generated within the particle rest frame
following the blast-wave description.  After generating
several thousands points, the number of relative positions ${\pmb r}_a - {\pmb r}_b$
for the relative source becomes of the order of a few millions, sufficient for finding quite
smooth source moments,
except at the lowest $|{\pmb r}_a - {\pmb r}_b|$.  We associate the $z$-axis direction with
the beam axis and the $x$-axis direction with the outward direction along the pair momentum.
The angular moments for the blast-wave model are compared in Fig.~\ref{fig:sr2_blast}
to the moments for a Gaussian source with the same cartesian moments, $\langle x\rangle$,
$\langle x^2 \rangle$, $\langle y^2 \rangle$ and $\langle z^2\rangle$, as the blast-wave
relative source.

As elsewhere, values for the underlying three-dimensional source structure can be restored
along any direction from the cartesian coefficients, in particular from those displayed
in Fig.~\ref{fig:sr2_blast},
exploiting tracelessness.  Thus e.g.\ along the $x$-axis the source
values are
\begin{equation}\label{eq:Sx}
  {\mathcal S} = {\mathcal S}^{(0)} \pm {\mathcal S}_x^{(1)} + \frac{1}{2}
  \left( S_{xx}^{(2)} - S_{yy}^{(2)} \right) - \frac{1}{2} S_{zz}^{(2)}
  \mp S_{xyy}^{(3)} \mp S_{xzz}^{(3)} + \ldots \, ,
\end{equation}
where the upper and lower signs refer to the positive and negative directions along the axis.
Within the $xy$-plane, at 45$^\circ$ to the $x$-axis, the source values are
\begin{equation}\label{eq:Sxy}
  {\mathcal S} = {\mathcal S}^{(0)} \pm \frac{1}{\sqrt{2}} \, {\mathcal S}_x^{(1)}
  - \frac{1}{2} \, S_{zz}^{(2)} \mp \frac{1}{2\sqrt{2}} \, S_{xzz}^{(3)} + \ldots \, ,
\end{equation}
with the upper signs again referring to the positive direction along the $x$-axis.

Within the blast-wave model, the heavier particles end up being emitted farther out,
on the average, along the outward direction, than the lighter particles.  This is due
to the following.  In an infinite system with Hubble expansion, a particle with a given velocity
would get, on the average, emitted
from a spatial patch centered around the point with the collective velocity matching the particle velocity.
The size of such an emission patch would depend on the particle mass, with the patch being tighter for heavier
particles with lower thermal velocities and being more extended for lighter particles.  Relative to
the infinite system, however, in a system with a finite transverse extension, the outer
portions of the emission patches would get cut off and more so for the light than for heavy particles, forcing
the emission region of the light particles inwards relative to the heavy.
In the past, relative displacements
of emission between particles of different mass, e.g.\ $K\pi$,
have been accessed by comparing the correlations
at opposite signs of outward momentum, $q_{\rm out}>0$ and $q_{\rm out}< 0$, at low values of
the remaining $q$-components~\cite{Adams:2003qa,Retiere:2002mb}.  Such a~strategy discards potentially
useful information
on the relative displacement in the remainder of the relative momentum space.
The displacement of species within a pair is visible in Fig.~\ref{fig:sr2_blast}
in the finite values of ${\mathcal S}^{(1)}_x$.  With the mass difference being
larger within the $p\pi$ than $pK$ pairs, the ${\mathcal S}^{(1)}_x$ coefficients
peak farther out and reach significantly higher values for the $p \pi$ than $pK$ pairs.

Some other features of blast-wave sources become apparent in the context of comparisons to gaussian sources.
When making such a comparison, a few points need to be kept in mind.
In an infinite system with a Hubble-like expansion along the longitudinal direction,
the associated relative source should fall off exponentially at large $r$ in the longitudinal direction,
rather than falling in a gaussian fashion, see e.g.\ Eq.~(\ref{eq:dNther}).
On the other hand, when the emission within a model gets confined to a finite transverse domain,
the source function should fall off more sharply in the transverse directions,
especially in the sideward direction which is less affected by decays.
For our comparison we choose the parameters of the  gaussian source to match the norm and
$\langle r_i \rangle$ and $\langle r_i \, r_j \rangle$ of the blast-wave source.
Given Eq.~(\ref{eq:cartesiansourcemoments}), this means that the $\ell \le 2$
integrals of $r^{\ell+2} \, {\mathcal S}^{(\ell)} (r)$ for the gaussian are made to
match those for the blast-wave.

When comparing the blast-wave and gaussian source-coefficients in Fig.~\ref{fig:sr2_blast},
we can see that an agreement between them generally worsens as $\ell$ increases and it is worse for $p \pi$
than $pK$ pairs.  These tendencies are to an extent associated with the source tails.  Since the emission
region region is wider in the case of pions than kaons,  the larger-scale geometry of the
emitting region is explored more stringently in the case of pions.  In addition, the
pions become relativistic at lower $p_t$ than do kaons.
The finite transverse extension of the emission zone most strongly suppresses the tail of the relative
source in the negative
$x$-direction.  In terms of angular coefficients, this suppression generally
results in a large enhancement of the
high-$\ell$ blast-wave coefficients at high-$r$, compared to the gaussian coefficients.  This
enhancement is particularly
visible for the $p\pi$ pairs.  The one case in Fig.~\ref{fig:sr2_blast}
where the $p\pi$ ${\mathcal S}_{zz}^{(2)}$ is smaller in magnitude
for the blast-wave than gaussian source is associated with the change in sign for that coefficient.
The change in sign is itself associated with the trimming of the relative source
by extension of the emitting region, present in the $x$ and $y$ directions, but absent in the $z$-direction.
Further sign changes are seen for the blast-wave $\ell=3$ coefficients, while there are no such
changes for
the corresponding gaussian coefficients.  These sign changes reflect the more complex nature
of the blast-wave sources of a crescent-like shape in the $xy$ and $xz$  directions.  The latter shape results from the relativistically noninstantaneous
nature of the boost-invariant blast-wave emission, mentioned already in the preceding section.
Coded in the ${\mathcal S}_{xyy}^{(3)}$ and ${\mathcal S}_{xzz}^{(3)}$ coefficients, cf.\
(\ref{eq:cartesiansourcemoments}), are the violations of the factorizations which hold for
the gaussian source: $\langle x \, y^2 \rangle = \langle x \rangle \, \langle y^2 \rangle$ and
$\langle x \, z^2 \rangle = \langle x \rangle \, \langle z^2 \rangle$. These violations result in significant differences between the $\ell=3$ coefficients for the gaussian and blast wave.

Regarding the examples of source coefficients in Fig.~\ref{fig:sr2_blast}, it is important to note that,
while the $\ell=0$ coefficients have most of their strength within the $(5-15) \, \mbox{fm}$ region
of separation,
the $\ell \ge 1$ coefficients have most of their strength within the $(7-25) \, \mbox{fm}$ region.
The fact that the source anisotropies are most pronounced at large~$r$, combined with the difference
in fall-off for strong and Coulomb kernels, of $1/r^2$ vs $1/r$, allows the Coulomb induced correlations
to compete effectively with the correlations induced by
strong interactions in providing access to the source anisotropies.

\begin{figure}
\centerline{\includegraphics[width=0.7\textwidth]{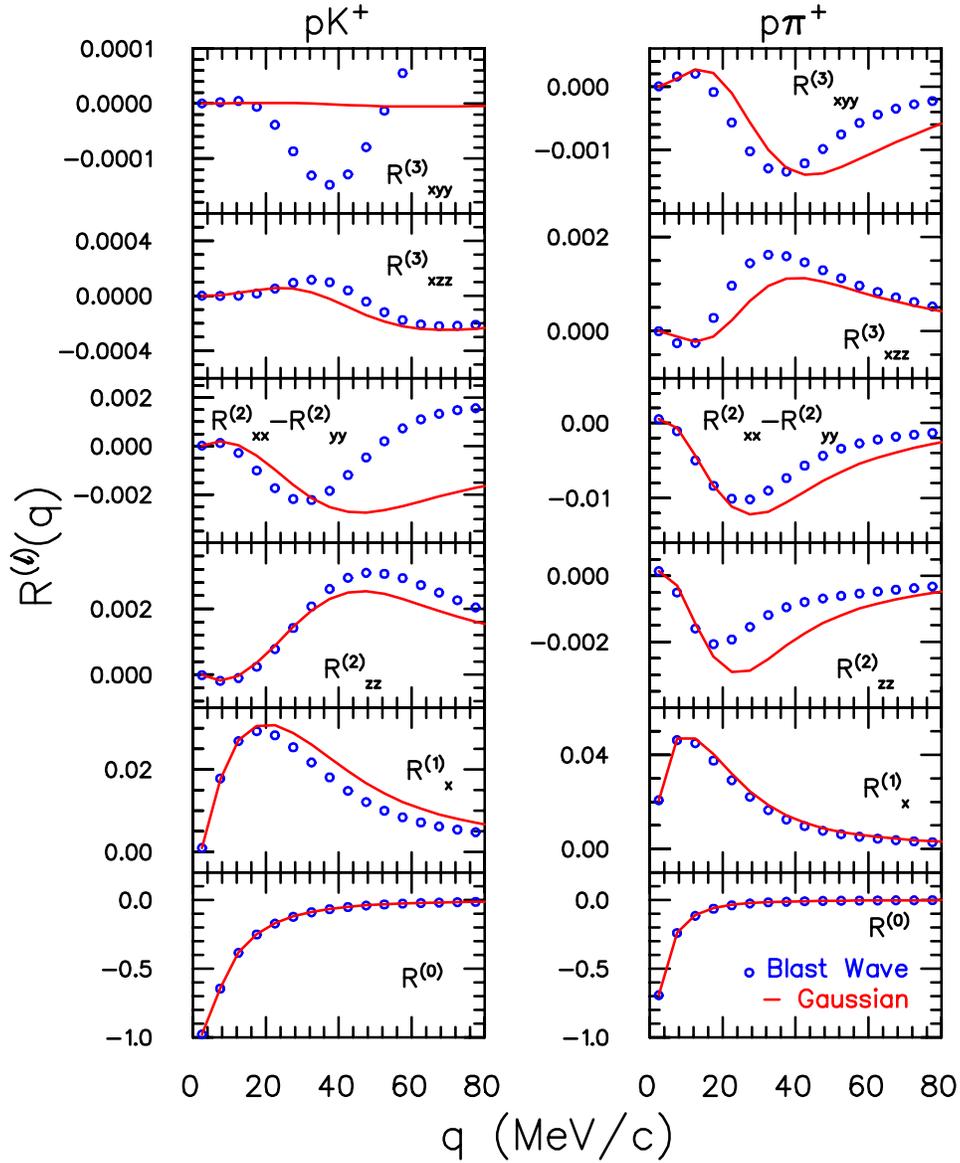}}
\caption{\label{fig:c_blast} (Color online) 
Cartesian coefficients for the correlations of $pK^+$ (left) and $p\pi^+$ (right) pairs
moving out at transverse velocity of $0.6 \, \mbox{c}$ relative to the beam axis.
Symbols represent the results obtained in the blast-wave model, while lines represent the results from
a gaussian source with the same pair c.m.\ cartesian moments, $\langle r_i\rangle$ and
$\langle r_i \, r_j\rangle$, as the blast-wave source.  The corresponding cartesian source
coefficients have been represented in Fig.~\ref{fig:sr2_blast}.  As in the case of
Fig.~\ref{fig:sr2_blast}, differences between the blast-wave and gaussian results are
more pronounced for the $p\pi^+$ than for $pK^+$ pairs.}
\end{figure}

Figure~\ref{fig:c_blast} displays the proton-charged meson correlations calculated by convoluting
the source functions illustrated in Fig.~\ref{fig:sr2_blast} with the respective kernels.
The utilized $pK^+$ and $p\pi^+$ kernels have been obtained accounting both for
the Coulomb as well as the strong interactions~\cite{Pratt:2003ar}.
Relative magnitudes of discrepancies between the blast-wave and gaussian are
similar for the corresponding correlation and source coefficients.  The level of
difficulty in accessing the shape information may be assessed in Fig.~\ref{fig:c_blast}
by examining the absolute magnitudes of coefficients for the different $\ell$-values.
For $\ell=1$, the magnitude is of the order of few percent, which should be observable
with statistics characteristic for the past correlation measurements.  For $\ell \ge 2$,
the magnitude is of the order of one percent and less.  Particularly well suited for investigating
the high-$\ell$ correlations should be the high-statistics
data from the 2004 run at RHIC.  Eventually, the reach of the analysis may be limited
by systematics rather than by statistics.

\section{Summary}
\label{sec:summary}

Correlation measurements are a crucial tool for
understanding the dynamics of central nuclear reactions.
One of the forefront tasks before the correlation analysis is to decipher the mire of RHIC data.
Without carrying out a multidimensional correlation analysis, there is little chance
for a detailed understanding of the impact of dynamics on correlations in those reactions and
for understanding of the role played by
the equation of state.
To date, correlation analyses at RHIC have focused on identical pions.
In identical-pion correlations, strong interactions play little role.  In the analyses to date, Coulomb interactions have
been normally corrected for in a shape-independent manner, leaving off correlations induced
by identical-particle interference.  The latter are related to the source through Fourier
transformation allowing for easy modelling with specific sources such as gaussian.  The decomposition
of correlations in terms of cartesian coefficients offers an alternative way of carrying out
a multidimensional correlation analysis for identical pion pairs and
it provides possibility for carrying out
multidimensional analysis for other classes of final-state effects.

One of the benefits of decomposing information with angular harmonics is that the
correlations induced by Coulomb and strong interactions carry, on their own,
some shape information that can be exploited in the correlation analysis.
The representation of the correlations in terms of the cartesian
or tesseral coefficients carries the full three-dimensional information in terms
of a set of one-dimensional functions.  Data represented in this fashion can be then
be compared to models visually, one harmonic at a time, providing the means to better identify
which shape characteristics of the data are being described by the model.

The ability to discern information about source shapes from correlation
measurements depends on the structure of the kernels, ${\mathcal
K}_\ell(q,r)$, for different spherical ranks~$\ell$.  In Section \ref{sec:kernels}, it was demonstrated that
the identical-particle interference, Coulomb and strong interactions
all produce effective kernels.
The resolving power of Coulomb interactions was shown to rise with mass and charge within the interacting
pair.  Strong interactions are most effective in the resonance region and, otherwise, for large cross sections.
Even purely $s$ wave interactions can produce kernels of good resolving power at higher $\ell$.

While expansions employing tesseral harmonics nominally provide as powerful means for storing information
in correlation functions as cartesian harmonics, they do not provide as transparent or as intuitive means
for understanding the shapes as expansions done using the cartesian harmonic basis.  Cartesian harmonics
have a variety of useful properties
which can be exploited in manipulating the angular
information as was shown in Sec.~\ref{sec:cartesian}.
The connection between specific aspects of the source geometry and specific angular moments
has been emphasized
in Sec.~\ref{sec:source}.  It has been pointed out that the average relative displacement of emission points in
the c.m.\ frame of nonidentical particles can be assessed through the $\ell=1$ angular moments.
The quadrupole distortion, most commonly quantified in terms of the radii $R_{\rm out}$, $R_{\rm side}$
and $R_{\rm long}$, can be studied with the $\ell=2$ moments.  The $\ell=3$ moments test the more
complex aspects of sources such as boomerang features.

Detailed examples of sources and correlations have been presented in Sec.~\ref{sec:examples}.
Using gaussian sources, it has been, in particular, demonstrated that the $p K^+$ and
$p \pi^+$ correlations can realistically be expected to provide shape information on emission sources.
The strengths of correlations characterizing shape anisotropies can reach
the magnitude of few percent.  While the upcoming RHIC data sets will provide sufficient statistics
to assess correlations between different particles at the level of one percent or better,
the true challenge may turn out to the subtraction of competing correlations, such as those
produced by jets.  The proton-charged meson correlations are dominated by Coulomb interactions.
The strong interactions in baryon-baryon systems can also, though, provide
good means for discerning shape anisotropies at low relative momenta, yielding
correlations of the same order or larger
than the proton-meson correlations.
One common feature of all pairs and sources examined here is that the variation of the correlation
with angle is comparable in magnitude to the angle-integrated correlation.
Consequently, if one has sufficient statistics to measure a one-dimensional correlation, only
a modest enhancement in statistics would suffice to find, or constrain in a meaningful manner,
the higher angular moments.  In the context of the blast-wave sources, opportunities for detecting
more complex shapes have been discussed, such as boomerang shapes.

In conclusion, a treasure trove of potential information lies largely ignored in the huge
volumes of data from high-energy collisions.  This information can address what
is the perhaps the greatest surprise of the first several years of RHIC experiments, the
source sizes and shapes inferred from identical-pion correlations.  The main
difficulty with the proposed new class of correlation analyses is the fact that they
rely on observables that are quantitatively small, of the order of a percent.
This will require
a careful analysis of competing correlations from jets and collective flow, but
we believe that the objectives will indeed prove feasible.

\appendix\section{Projection Operator ${\mathcal P}$}

In arriving at the cartesian harmonics, an important role has been played by
the cartesian projection operator ${\mathcal P}$ which projects out the
symmetric traceless portion of any cartesian tensor.  As a projection operator,
the operator ${\mathcal P}$ must have the following properties:
\begin{eqnarray}
\label{eq:Ptrans}
{\mathcal P}_{\alpha_1 \ldots \alpha_\ell : \alpha_1' \ldots \alpha_\ell' }^{(\ell:\ell)}
& = &
{\mathcal P}_{\alpha_1' \ldots \alpha_\ell' : \alpha_1 \ldots \alpha_\ell }^{(\ell:\ell)} \\
{\mathcal P}_{\alpha_1 \ldots \alpha_\ell : \alpha_1' \ldots \alpha_\ell' }^{(\ell:\ell)}
& = & \sum_{\alpha_1" \ldots \alpha_\ell" }
{\mathcal P}_{\alpha_1 \ldots \alpha_\ell : \alpha_1" \ldots \alpha_\ell" }^{(\ell:\ell)}
\,
{\mathcal P}_{\alpha_1" \ldots \alpha_\ell" : \alpha_1' \ldots \alpha_\ell' }^{(\ell:\ell)}
\, .
\end{eqnarray}
The projection operator acting on any tensor $T$ must yield a traceless result
${\mathcal P}T$.  Since we can choose, in particular, a tensor that is finite
for any single set of indices and zero otherwise, we find that the operator
${\mathcal P}$ itself must be traceless:
\begin{equation}\label{eq:Ptrace1}
  \sum_{\alpha } {\mathcal P}_{\alpha_1 \ldots \alpha_{\ell-2} \, \alpha \,
  \alpha : \alpha_1' \ldots \alpha_\ell' }^{(\ell:\ell)}
  =
  \sum_{\alpha  }
  {\mathcal P}_{\alpha_1 \ldots
  \alpha_\ell : \alpha_1' \ldots \alpha_{\ell-2}' \, \alpha \, \alpha }^{(\ell:\ell)}
 = 0 \, .
\end{equation}
We shall demonstrate existence of the operator ${\mathcal P}$ by construction.
Notably, when acting onto a symmetrized traceless tensor, the operator must
reproduce the tensor.  Thus, we can represent ${\mathcal P}$ as
\begin{equation}\label{eq:PSd}
 {\mathcal P}_{\alpha_1 \ldots \alpha_\ell : \alpha_1' \ldots \alpha_\ell' }^{(\ell:\ell)}
 = {S}_{\alpha_1 \ldots \alpha_\ell : \alpha_1' \ldots \alpha_\ell' }^{(\ell:\ell)}
 +
 \Delta {\mathcal P}_{\alpha_1 \ldots \alpha_\ell : \alpha_1' \ldots \alpha_\ell' }^{(\ell:\ell)} \, ,
\end{equation}
or, in the shortened notation,
\begin{equation}\label{eq:PSdshort}
 {\mathcal P}^{(\ell:\ell)}
 = { S}^{(\ell:\ell)}
 +
 \Delta {\mathcal P}^{(\ell:\ell)} \, ,
\end{equation}
where ${S}$ is the symmetrized identity operator
\begin{equation}\label{eq:S=}
{ S}_{\alpha_1 \ldots \alpha_\ell : \alpha_1' \ldots \alpha_\ell' }^{(\ell:\ell)}
= \frac{1}{\ell !} \sum_{\sigma \in \Pi (\ell)} \delta_{\alpha_{\sigma (1)} \, \alpha_{1}' } \ldots
\delta_{\alpha_{\sigma(\ell)} \, \alpha_{\ell}' } \, ,
\end{equation}
and $\Delta {\mathcal P}$ is an operator which yields a zero when acting upon a
traceless tensor.  In (\ref{eq:S=}), the sum is over permutations of $\ell$
indices.  Importantly, a traceless symmetric tensor of rank $\ell '$ is
traceless and symmetric within any subset $\ell < \ell'$ of its indices.  When
acting on any of $\ell$ indices of a rank-$\ell'$, $\ell' > \ell$, traceless
tensor, the constructed operator ${\mathcal P}^{(\ell:\ell)}$ will reproduce
that tensor and $\Delta {\mathcal P}^{(\ell:\ell)}$ alone will annihilate it.

\begin{figure}
\centerline{\includegraphics[width=0.8\textwidth]{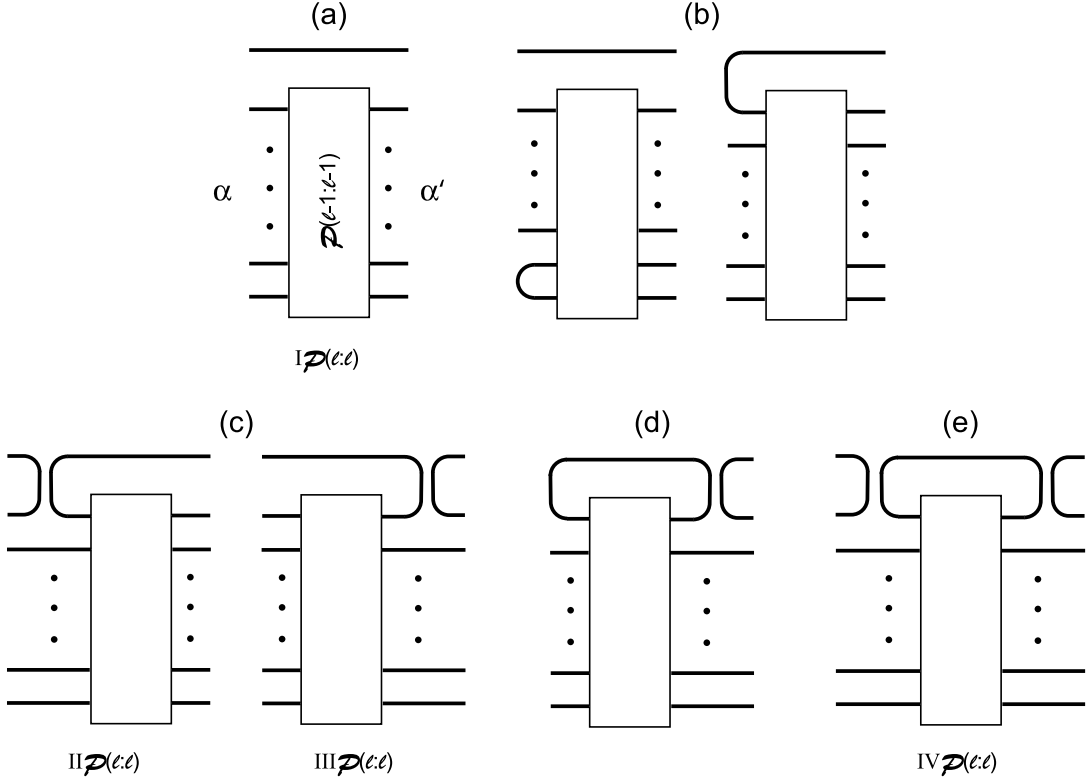}}
\caption{\label{fig:Pll} Diagrams illustrating the construction of the
  projection operator ${\mathcal P}^{(\ell:\ell)}$, using the lower-rank
  operator ${\mathcal P}^{(\ell-1:\ell-1)}$, represented by the rectangular
  block, and $\delta$-symbols, represented by lines.  (a)~Diagrammatic
  representation for the leading term $^I{\mathcal P}^{(\ell:\ell)}$,
  Eq.~(\protect\ref{eq:PI}).  (b)~Diagrammatic representation for the terms
  resulting from evaluating the trace of $^I{\mathcal P}^{(\ell:\ell)}$.
  The~terms represented by the left diagram yield zero due to the tracelessness
  of ${\mathcal P}^{(\ell-1:\ell-1)}$. (c)~Diagrammatic representation for the
  $^{II}{\mathcal P}^{(\ell:\ell)}$ and $^{III}{\mathcal P}^{(\ell:\ell)}$
  terms within the construct, Eqs.~(\protect\ref{eq:PII})
  and~(\protect\ref{eq:PIII}), respectively.  (d)~Representation for the
  nonvanishing terms from evaluating the trace of $^{III}{\mathcal
    P}$. (e)~Diagrammatic representation for the final $^{IV}{\mathcal
    P}^{(\ell:\ell)}$ term within the construct, Eq.~(\protect\ref{eq:PIV}).  }
\end{figure}

In our construction of ${\mathcal P}$, we will find a graphical method similar
to that employed for interactions~\cite{Cvitanovic:1976am} useful.  Within the
method, the Kronecker $\delta_{\alpha \, \alpha'}$ is represented as a line
joining the indices $\alpha$ and $\alpha '$.  A convolution of delta symbols
$\sum_{\alpha_1 \ldots \alpha_n} \delta_{\alpha \, \alpha_1} \,
\delta_{\alpha_1 \, \alpha_2} \ldots \delta_{\alpha_n \, \alpha'} =
\delta_{\alpha \, \alpha'}$ is represented by joining of the line segments
which produces a single line from $\alpha$ to $\alpha'$.  A line which closes
on itself represents a scalar factor which is the trace of the $\delta$-symbol,
equal to the dimension of space, $\sum_\alpha \delta_{\alpha \, \alpha}=3$.
Examples will be provided in Fig.~\ref{fig:Pll}.

For $\ell =0 $ and $\ell = 1$, the operator ${\mathcal P}$ is simply the
identity operator, ${\mathcal P}^{(0:0)}=1$ and ${\mathcal P}_{\alpha : \alpha
  '}^{(1:1)}=\delta_{\alpha \, \alpha '}$.  We shall show that, when the
operator for the rank $(\ell -1)$ tensors is known, it is possible to construct
the operator with the required properties for the next rank $\ell$ tensors.
Thus, the operator for any rank can be constructed by recursion.  Let us assume
that we know the operator for rank $(\ell-1)$, ${\mathcal P}^{(\ell -1 : \ell
  -1)}$.  We shall construct the next rank operator as a sum of terms,
\begin{equation}\label{eq:Pconst}
  {\mathcal P}^{(\ell;\ell)} =  {^I {\mathcal P}}^{(\ell;\ell)} + {^{II} {\mathcal P}}^{(\ell;\ell)}
  + \ldots \, ,
\end{equation}
exploiting the operator ${\mathcal P}^{(\ell -1 : \ell - 1)}$.  The terms of
different structure will be added to achieve in (\ref{eq:Pconst}) the
\mbox{properties~(\ref{eq:Ptrans})-(\ref{eq:Ptrace1})} for ${\mathcal P}^{(\ell
  : \ell)}$.  In particular, only $^I{\mathcal P}$ in our construction will
contain the symmetrization operator $S$ on the r.h.s.\ of
Eq.~(\ref{eq:PSdshort}).

Since, within each rank, the operator must, on one hand, contain a symmetrized
identity operator and, on the other, be traceless, it is natural to start with
${^I {\mathcal P}}$ as a tensor product of the lower-rank operator and
identity, illustrated in Fig.~\ref{fig:Pll}(a),
\begin{equation}\label{eq:PI}
  \delta_{\alpha_{\ell} \, \alpha_{\ell}'}
  \,
  {\mathcal P}^{(\ell-1;\ell-1)}_{\alpha_1 \ldots \alpha_{\ell -1 }:\alpha_1' \ldots \alpha_{\ell -1}'}
  \, ,
\end{equation}
symmetrized separately in the index sets $\alpha$ and $\alpha '$.
Evaluation of the trace for $^I {\mathcal P}$, produces terms which correspond
to the trace of the lower-rank
operator ${\mathcal P}^{(\ell -1 : \ell - 1)}$, represented by the left
diagram in Fig.~\ref{fig:Pll}(b), and to the convolutions of
the $\delta$-symbol with ${\mathcal P}^{(\ell -1 : \ell - 1)}$,
represented by right diagram in Fig.~\ref{fig:Pll}(b).  Terms of
the first type yield zero due to the tracelessness of
${\mathcal P}^{(\ell -1 : \ell - 1)}$, but those of the second
type generally yield a finite contribution.  Aiming at the
tracelessness of ${\mathcal P}^{(\ell : \ell )}$, one can next add to the construct a counter-term
$^{II} {\mathcal P}$ of the form
\begin{equation}\label{eq:PII}
- {c_{II}} \,
\delta_{\alpha_{\ell - 1} \, \alpha_{\ell}}
\,
{\mathcal P}^{(\ell - 1;\ell -1)}_{\alpha_1 \ldots \alpha_{\ell -2}
\, \alpha_{\ell}' :\alpha_1' \ldots \alpha_{\ell -1}'}
\, ,
\end{equation}
to cancel the last contribution, symmetrized separately in the indices $\alpha$ and $\alpha '$,
represented by the left diagram in
Fig.~\ref{fig:Pll}(c).  The constant ${c_{II}}$ should be adjusted to achieve the tracelessness of
${\mathcal P}^{(\ell :\ell )}$.  Taking the trace of $^{II} {\mathcal P}$ produces the same type of terms
as for $^{II} {\mathcal P}$, that were
represented in Fig.~\ref{fig:Pll}(b).  Thus, an adjustment of ${c_{II}}$ can indeed produce a
traceless sum $^{I} {\mathcal P}+{^{II} {\mathcal P}}$.
However, the operator ${\mathcal P}^{(\ell :\ell )}$
should be symmetric under the transposition of the indices $\alpha$ and $\alpha'$, Eq.~(\ref{eq:Ptrans}).
In~consequence, when adding the term $^{II} {\mathcal P}$ to the construction, one also needs to add a term
${^{III} {\mathcal P}} = \left({^{II} {\mathcal P}}\right)^\top$, of the form
\begin{equation}\label{eq:PIII}
- {c_{II}} \,
\delta_{\alpha_{\ell -1}' \, \alpha_{\ell}'}
\,
{\mathcal P}^{(\ell -1;\ell -1)}_{\alpha_1 \ldots \alpha_{\ell-1} : \alpha_1' \ldots \alpha_{\ell -2}'
\, \alpha_{\ell} }
\, ,
\end{equation}
symmetrized separately in the index sets $\alpha$ and $\alpha '$ and represented by the right diagram in
Fig.~\ref{fig:Pll}(c).  The nonvanishing terms from evaluating the trace of $^{III} {\mathcal P}$ are
illustrated with the diagram in Fig.~\ref{fig:Pll}(d).  That last contribution to the trace may be compensated
for by adding the final term $^{IV} {\mathcal P}$ to the construct, symmetric under transposition, of the form
\begin{equation}\label{eq:PIV}
c_{IV}
\, \delta_{\alpha_{\ell - 1} \, \alpha_{\ell }}
\, \delta_{\alpha_{\ell - 1}' \, \alpha_{\ell}'}
\sum_\beta
{\mathcal P}^{(\ell -1;\ell - 1)}_{\alpha_1 \ldots \alpha_{\ell -2} \, \beta
: \alpha_1' \ldots \alpha_{\ell -2}' \, \beta }
\, ,
\end{equation}
symmetrized separately in the indices $\alpha$ and $\alpha '$, and illustrated with the diagram in
Fig.~\ref{fig:Pll}(d).  Taking the trace of $^{IV} {\mathcal P}$ produces, up to a factor, the same result as
taking the trace of $^{III} {\mathcal P}$.

The constants $c_{II}$ and $c_{IV}$ can be now set to make the trace of ${\mathcal P}^{(\ell ;\ell )}$
vanish.  The net result for the operator, with the symmetrizations made explicit, is
\begin{eqnarray}
{\mathcal P}^{(\ell ;\ell )}_{\alpha_1 \ldots \alpha_{\ell }:\alpha_1' \ldots \alpha_{\ell }'}
& =
& \frac{1}{\ell^2} \sum_{m,n =1}^{\ell}
\delta_{\alpha_m \, \alpha_n'}
\,
 {\mathcal P}_{\alpha_1 \ldots \alpha_{m-1} \,
\alpha_{m+1} \ldots \alpha_{\ell} : \alpha_1' \ldots \alpha_{n-1}' \,
\alpha_{n+1}' \ldots \alpha_{\ell}'}^{(\ell -1:\ell -1)}
\nonumber \\
&& - \frac{2}{\ell^2 \, (2 \ell -1)} \sum_{\substack{1 \le m < r \le \ell \\ 1 \le n \le \ell }}
\delta_{\alpha_m \, \alpha_r}
\,
{\mathcal P}_{\alpha_1 \ldots \alpha_{m-1} \,
\alpha_{m+1} \ldots \alpha_{r - 1} \, \alpha_{r+1} \ldots
\alpha_{\ell} \, \alpha_n': \alpha_1' \ldots \alpha_{n-1}' \,
\alpha_{n+1}' \ldots \alpha_{\ell}'}^{(\ell -1:\ell -1)}
\nonumber \\
&& - \frac{2}{\ell^2 \, (2 \ell -1)} \sum_{\substack{1 \le m  \le \ell \\ 1 \le n < s \le \ell }}
\delta_{\alpha_n' \, \alpha_s'}
\,
{\mathcal P}_{\alpha_1 \ldots \alpha_{m-1} \,
\alpha_{m+1} \ldots
\alpha_{\ell}: \alpha_1' \ldots \alpha_{n-1}' \, \alpha_{n+1}'  \ldots \alpha_{s - 1}' \, \alpha_{s+1}'
\ldots \alpha_{\ell }'  \, \alpha_m}^{(\ell -1:\ell -1)}
\nonumber \\
&& + \frac{4}{\ell^2 \, (2 \ell -1)^2} \sum_{\substack{1 \le m < r \le \ell \\ 1 \le n < s \le \ell }}
\delta_{\alpha_m \, \alpha_r}
\,
\delta_{\alpha_n' \, \alpha_s'}
\nonumber
\\
&&  \times
\sum_\beta
{\mathcal P}_{\alpha_1 \ldots \alpha_{m-1} \,
\alpha_{m+1} \ldots \alpha_{r - 1} \, \alpha_{r+1} \ldots
\alpha_{\ell} \, \beta:
\alpha_1' \ldots \alpha_{n-1}' \, \alpha_{n+1}'  \ldots \alpha_{s - 1}' \, \alpha_{s+1}'
\ldots \alpha_{\ell}'  \, \beta }^{(\ell -1:\ell -1)} \, .
\label{eq:Pll1}
\end{eqnarray}
By now, we have constructed an operator ${\mathcal P}^{(\ell ;\ell )}$ which is symmetric separately in the
covariant and contravariant indices, traceless and symmetric under the transposition.
The remaining question is whether repeated applications of the operator yield the same result, i.e.\
whether Eq.~(\ref{eq:Ptrans}) is satisfied.

To answer the last question,
let us apply ${\mathcal P}^{(\ell ;\ell )}$,
out of the four components in Eqs.~(\ref{eq:Pconst}) and (\ref{eq:Pll1}),
to the constructed
${\mathcal P}^{(\ell ;\ell )}$ with the properties we have established.
The $^I{\mathcal P}$ component reproduces
${\mathcal P}^{(\ell ;\ell )}$, as both the Kronecker symbol and ${\mathcal P}^{(\ell -1;\ell -1)}$
reproduce a
traceless symmetric tensor.  On the other hand, all the other components of the construct annihilate
${\mathcal P}^{(\ell ;\ell )}$, because they involve calculation of the trace of that operator, either
directly or after application of ${\mathcal P}^{(\ell -1;\ell -1)}$.
Thus, Eq.~(\ref{eq:Ptrans}) is satisfied.

Use of Eq.~(\ref{eq:Pll1}) for $\ell=2$ produces
\begin{equation}\label{eq:P2}
{\mathcal P}_{\alpha_1 \, \alpha_2 : \alpha_1' \, \alpha_2'}^{(2:2)}
= \frac{1}{2}
\left(
\delta_{\alpha_1 \, \alpha_1'} \, \delta_{\alpha_2 \, \alpha_2'}
+
\delta_{\alpha_1 \, \alpha_2'} \, \delta_{\alpha_2 \, \alpha_1'}
\right)
- \frac{1}{3} \,
\delta_{\alpha_1 \, \alpha_2} \, \delta_{\alpha_1' \, \alpha_2'}
\, .
\end{equation}
In general, for ${\mathcal P}^{(\ell;\ell)}$ at $\ell \ge 2$,
the repeated recursion with Eq.~(\ref{eq:Pll1}) produces,
besides the leading symmetrized identity
term, the correction terms which involve replacing $k$ pairs of Kronecker symbols in the identity, linking
the covariant and contravariant indices, with the pairs of Kronecker symbols linking separately
the covariant and contravariant
indices, $k=1, 2 \ldots \ell/2$.
Application of both sides of Eq.~(\ref{eq:Pll1}) for ${\mathcal P}$ to
${\pmb n}^{\ell}$ produces the recursion relation
(\ref{eq:Arecursion}) for~${\mathcal A}$.

\begin{acknowledgments}
The authors acknowledge insightful discussions with Paul Chung, David Brown and Richard Lednicky.
Support was provided by the U.S.\ National Science Foundation, Grant No.\ PHY-0555893,
and by the U.S.\ Department of Energy, Grant No.\ DE-FG02-03ER41259.
\end{acknowledgments}


\end{document}